\documentclass[twocolumn,superscriptaddress,nofootinbib,amsmath,amssymb,amsfonts]{aastex631}

\usepackage{newtxtext,newtxmath}

\usepackage{graphicx}%
\usepackage{dcolumn}%
\usepackage{enumitem}%
\usepackage{bm}%
\usepackage{hyperref}%
\hypersetup{
  colorlinks=true,
  linkcolor=blue,
  citecolor=cyan,
  urlcolor=cyan
}
\usepackage{colortbl}
\usepackage{amsmath,amssymb,amsfonts}
\usepackage{times}

\newcommand{\ie}{i.e.,~}
\newcommand{\eg}{e.g.,~}

\begin{document}

\title{On the impact of neutrinos on the launching of relativistic jets
  from ``magnetars'' produced in neutron-star mergers}

\author[0000-0002-9955-3451]{Carlo Musolino}
\affiliation{Institut f\"ur Theoretische Physik, Goethe Universit\"at, Max-von-Laue-Stra{\ss}e 1, 60438 Frankfurt am Main, Germany}
\author[0000-0002-1330-7103]{Luciano Rezzolla}
\affiliation{Institut f\"ur Theoretische Physik, Goethe Universit\"at, Max-von-Laue-Stra{\ss}e 1, 60438 Frankfurt am Main, Germany}
\affiliation{School of Mathematics, Trinity College, Dublin 2, Ireland}
\affiliation{Frankfurt Institute for Advanced Studies, Ruth-Moufang-Str. 1, 60438 Frankfurt am Main, Germany}
\author[0000-0002-0491-1210]{Elias R. Most}
\affiliation{TAPIR, Mailcode 350-17, California Institute of Technology, 1200 E California Blvd, Pasadena, CA 91125, USA}
\affiliation{Walter Burke Institute for Theoretical Physics, California Institute of Technology, 1200 E California Blvd, Pasadena, CA 91125, USA}

\date{\today}

\begin{abstract}
A significant interest has emerged recently in assessing whether
collimated and ultra-relativistic outflows can be produced by a
long-lived remnant from a binary neutron-star (BNS) merger, with
different approaches leading to different outcomes. To clarify some of
the aspect of this process, we report the results of long-term (\ie
$\sim~110\,{\rm ms}$) state-of-the-art general-relativistic
magnetohydrodynamics simulations of the inspiral and merger of a BNS
system of magnetized stars. We find that after $\sim~50\,{\rm ms}$ from
the merger, an $\alpha$-$\Omega$~dynamo driven by the magnetorotational
instability (MRI) sets-in in the densest regions of the disk and leads to
the breakout of the magnetic-field lines from the accretion disk around
the remnant. The breakout, which can be associated with the violation of
the Parker-stability criterion, is responsible for the generation of a
collimated, magnetically-driven outflow with only mildly relativistic
velocities that is responsible for a violent eruption of electromagnetic
energy. We provide evidence that this outflow is partly collimated via a
Blandford-Payne mechanism driven by the open field lines anchored in the
inner disk regions. Finally, by including or not the radiative transport
via neutrinos, we determine the role they play in the launching of the
collimated wind. In this way, we conclude that the mechanism of
magnetic-field breakout we observe is robust and takes place even without
neutrinos. Contrary to previous expectations, the inclusion of neutrinos
absorption and emission leads to a smaller baryon pollution in polar
regions, and hence accelerates the occurrence of the breakout, yielding a
larger electromagnetic luminosity. Given the mildly relativistic nature
of these disk-driven breakout outflows, it is difficult to consider them
responsible for the jet phenomenology observed in short gamma-ray bursts.
\end{abstract}


\section{Introduction}

Binary neutron star (BNS) mergers are known to be the progenitors of
transient electromagnetic (EM) emission phenomena such as kilonovae and
short gamma-ray bursts. The mass distribution of pulsars in galactic
binaries, combined with estimates on the maximum mass that can be
supported against gravitational collapse by the equation of state (EOS),
and the results of numerical-relativity simulations, suggest that the
outcome of a typical merger of two neutron stars is a differentially
rotating hypermassive neutron star (HMNS), which survives for a timescale
on the order of $\sim~100\,{\rm ms}$ to $\sim~1\,{\rm s}$ before
collapsing to form a black hole~\citep[see, \eg][for some
  reviews]{Baiotti2016, Paschalidis2016}. This timescale is also relevant
for the development of the EM counterparts to BNS mergers, as testified
by the GW170817 event~\citep{Abbott2017_etal}, where a delay of
$\sim~1.7\,{\rm s}$ was measured between the EM and the
gravitational-wave (GW) signal. This time is normally associated to the
survival of the HMNS and the launching of a highly relativistic
collimated jet, which successfully broke out of the ejected material
surrounding the central engine~\citep{Gill2019, Murguia-Berthier2016,
  Murguia-Berthier2020}, see also \citet{Metzger2014} for additional
constraints from changes in the afterglow.

Since the launching of such a jet is likely to require an ordered and
large-scale magnetic field, it is important to study in detail whether
and how such a structure can emerge in the presence of a magnetized HMNS,
or ``magnetar''. Moreover, the massive blue kilonova that was also
observed following GW170817 implies that a large quantity of high
electron-fraction material, \ie with $Y_e > 0.25$, was ejected at mildly
relativistic speeds by the merger remnant. Since the progenitors in the
BNS system are initially very neutron rich, this material can only be
explained by the re-processing of the ejecta by weak interactions and
hence the irradiation of neutrinos \citep{Metzger2014}. Furthermore,
because the disk around the black hole formed after merger is unlikely to
reach temperatures and densities high enough to emit a sufficient amount
of electron anti-neutrinos, a long-lived HMNS remnant is probably the
only viable explanation to the blue-kilonova mass observed in
GW170817. Clearly, a detailed understanding of the evolution of a
magnetized HMNS on secular timescales is essential for the modelling of
the EM counterparts of BNS mergers in general, and for the launching of
collimated outflows from metastable BNS merger remnants, in particular.

A recent study based on long-term, numerical-relativity simulation of
black-hole--neutron-star mergers~\citep{Gottlieb:2023a} suggests that
while a sub-population of compact merger long gamma-ray bursts are likely
powered by black holes surrounded by an accretion disk, short gamma-ray
bursts may be better explained by magnetar engines
\citep{Gottlieb:2023b}. However, direct ab-initio numerical evidence of
whether a HMNS can power a gamma-ray burst remains
inconclusive~\citep[see, \eg][]{Ciolfi2020_a, Moesta2020, Most2023,
  Combi2023, Kiuchi2023, Bamber2024, Aguilera-Miret2024}. This is due,
in great part, to the significant challenges involved with simulating a
differentially rotating HMNS remnant at a sufficiently high resolution
and over a sufficiently long timescale to be able to track the evolution
of the topology and strength of the magnetic field.

By employing ultra-high resolutions, a recent work~\citep{Kiuchi2023} has
also highlighted how a magneto-rotational instability (MRI) driven
$\alpha$-$\Omega$~dynamo~\citep{Ruediger1993, Bonanno:2003uw}, might be
at play in the outer layers of the remnant, which might ultimately lead
to breakout of the magnetic field from the surface of the star and to a
magnetically driven collimated outflow. These direct simulations have
been followed by other studies that have tried to reproduce the effect of
an $\alpha$-$\Omega$~dynamo in simulations by incorporating effective
dynamo terms in the equations themselves~\citep{Most2023b}. An important
caveat in this process is the impact of neutrino-driven winds from the
stellar surface, which are expected to lower the Lorentz factor
reached in these outflows \citep{Dessart2009}.

The study presented here is meant to address and hopefully shed light on
a number of the open questions discussed above. In particular, we focus
on the impact of neutrinos on the outflows and present a launching
mechanism driven by buoyant magnetic field breakout from the remnant
accretion disk. To this end, we perform long-term and full
general-relativistic magnetohydrodynamics (GRMHD) evolutions of the
inspiral and merger of a binary system of magnetized neutron stars with a
full moment-based treatment of neutrinos. We find that contrary to
previous expectations, the inclusion of neutrinos do not lead to
additional baryon pollution of the outflows, nor do they inhibit magnetic
field breakout, but speed-up this process compared to a scenario where
they are not included.

\section{Methods}

We perform our simulations within the ideal-GRMHD
approximation~\citep[see, \eg][for a recent review]{Mizuno2024} with
adaptive-mesh-refinement (AMR) provided by the~\texttt{EinsteinToolkit}
\citep{EinsteinToolkit_etal:2020_11}. The GRMHD equations are solved with
fourth-order accurate finite differences by the
\texttt{Frankfurt/IllinoisGRMHD (FIL)} code~\citep{Most2019b,Etienne2015}
and the divergence-free constraint of the magnetic field is ensured via
constrained transport evolution of the magnetic vector potential
\citep{Londrillo2004} (see also \citet{Etienne:2010ui,Etienne2012a}).
\texttt{FIL} provides its own spacetime evolution code based on the Z4c
system \citep{Hilditch2012}, as well as a framework for handling
temperature and composition-dependent equations of state (EOSs)
\citep{Most:2018eaw,Most:2019onn}. Neutrinos are accounted for through
the energy-integrated, moment-based M1 scheme
\texttt{FIL-M1}~\citep[see][for details on our
  implementation]{Musolino2023}, which self-consistently accounts for
energy and momentum transfers between neutrinos and the astrophysical
plasma as well as for composition changes in the fluid due to weak
processes. In order to investigate the impact of dynamo processes in the
disk unambiguously, we do not employ the mean-field dynamo module
available in \texttt{FIL} \citep{Most2023b}. Last but not least, high
quality initial data for our binaries is obtained with the \texttt{FUKA}
solver~\cite{Papenfort2021b, Tootle2024a}.

Using this computational infrastructure, we have performed two long-term,
\ie $\gtrsim 110\,{\rm ms}$, simulations of equal-mass BNS mergers
described by the BHB$\Lambda\phi$ EOS \citep{Banik2014}, where each of
the star has an isolated ADM mass of $M_{\rm ADM}=1.25 \, M_\odot$. This
value is chosen as to ensure that the HMNS is metastable over a timescale
of a few $100\,{\rm ms}$ before collapsing to form a black hole. We
perform our simulations on a fixed mesh-refinement (FMR) grid with eight
levels, where the finest one has a spacing of $dx_{\rm f}\sim 210\,{\rm
  m}$ and covers a volume of $[-70, 70]^2 \times [0, 35]\,{\rm km}$ to
ensure that the HMNS and the densest regions of the disk are contained
within the finest refinement level. The entire domain covers $\sim
[-10^4, 10^4]^2 \times [0, 10^4]\,{\rm km}$ and we employ $z$-reflection
symmetry to save computational costs.

\begin{figure*}[t]
  \center
  \includegraphics[width=0.95\textwidth]{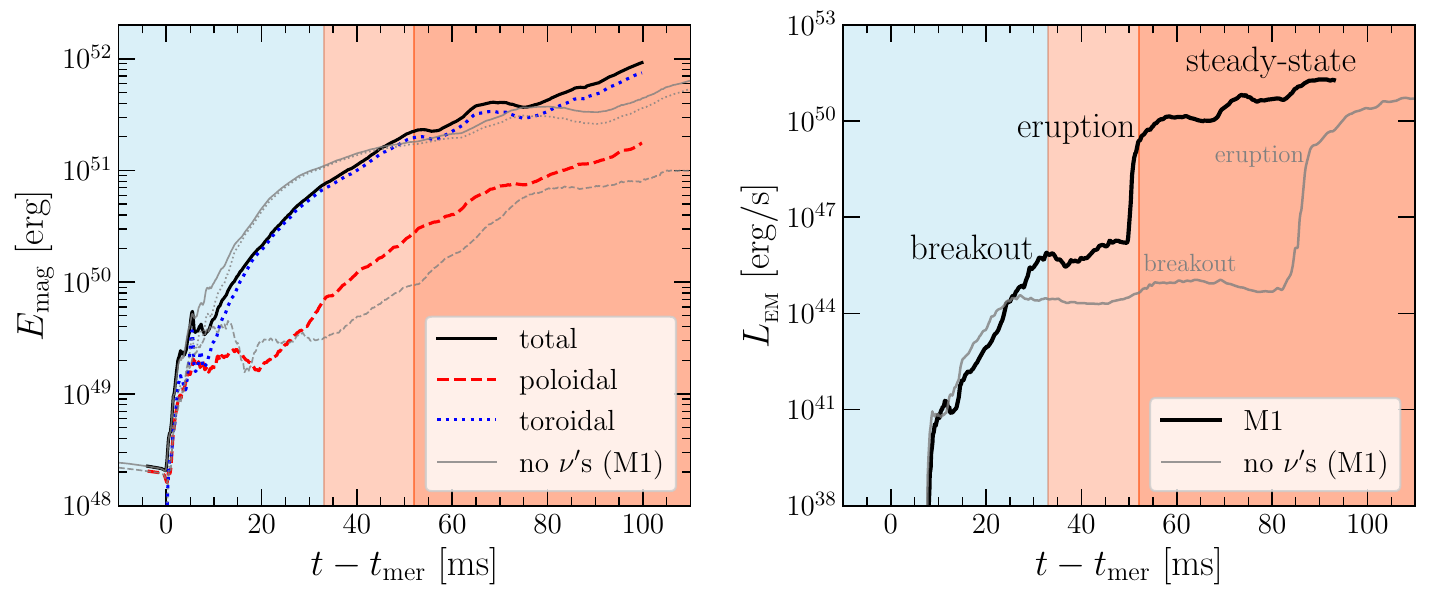}
  \caption{\textit{Left panel:} Evolution of the total magnetic energy
    (black solid line) but also of the poloidal (red dashed line) and
    toroidal components (blue dotted line). Also shown with gray lines of
    the same type are the corresponding evolutions when no neutrinos are
    taken into account (no ${\rm M1}$). \textit{Right panel:} Evolution
    of the EM luminosity during the simulation. Note the large jump when
    the magnetically-driven wind, that is emitted at the ``breakout''
    reaches the detector at the ``eruption'' time and is sustained in a
    steady state. Shown with gray lines is the luminosity in the absence
    of neutrinos (no ${\rm M1}$) and highlighting that neutrinos are
  useful to trigger the breakout but not necessary.}
  \label{fig:mag_obs}
\end{figure*}

The magnetic field is initially confined to the stars and is initialised
so that the maximum magnetic-field strength is $B_0=10^{16.5}\,{\rm
  G}$~\citep[see][for more details]{Chabanov2022}. We caution that our
choice of spatial resolutions -- dictated by the of acceptable
computational costs over the long timescales considered here -- is
insufficient to capture all aspects of magnetic field amplification
associated with the MRI-driven dynamo, which especially inside the outer
layers of the HMNS have been shown to require at least $70\,\rm m$
\citep{Kiuchi2017}. However, the mechanism presented in this work only
relies on dynamo processes in the accretion disk, which are fully
captured at this resolution. In addition, the extremely large value of
the magnetic field is chosen to be able to resolve at least partially the
MRI in the inner layers of the disk after merger. Finally, to assess the
role neutrinos play in the breakout of the magnetic field, and the baryon
pollution of polar outflows, we perform two sets of simulations
\textit{with and without} the inclusion of neutrinos vi an M1
scheme. This choice doubles our computational costs but, as we will
discuss, provides precious information.

\section{Results}

\subsection{Magnetic evolution and observables}

Because much of our focus here is on the long-term post-merger dynamics
of the remnant, we will not discuss in detail the main aspects of the
inspiral of magnetized binaries, which have been presented in several
works before~\citep[see, \eg][for some early investigations]{Liu:2008xy,
  Giacomazzo:2009mp, Palenzuela2015}. Similarly, we will not present the
complex dynamics that takes place at the merger and though the
development of the KHI and that has also been discussed in a number of
publications over the years~\citep[see, \eg][for some early and recent
  works]{Price06, Giacomazzo:2010, Kiuchi2015a, Aguilera-Miret2020,
  Chabanov2022}. Hence, our discussion will be restricted to the
post-merger remnant starting from a timescale of few tens of milliseconds
after the merger. At this point, the HMNS has reached a quasi-stationary
equilibrium that is essentially axisymmetric but for the presence of
small non-axisymmetric perturbations -- quantifiable mostly in the $m=1$
and $m=2$ modes~\citep[see, \eg][]{Lehner2016, Radice2016a, East2019,
  Papenfort:2022ywx, Topolski2024} -- and in a differential-rotation
profile characterized by a slowly uniformly rotating core and a Keplerian
mantle or ``disk''~\citep{Kastaun2014, Hanauske2016,Uryu2017, Cassing2024}.

Under these conditions, turbulence in the remnant is fully developed
(with properties that depend on resolution
\cite{Aguilera-Miret2021,Palenzuela_2022PRD, Chabanov2023}) and more
stationary, so that large-scale shearing motions can be produced and lead
to the so-called ``magnetic-winding''. We recall that, under the
infinite-conductivity conditions of ideal MHD, coherent large-scale
shearing motions lead a growth of the magnetic-field strength that is
linear in time \citep{Shapiro04}. Obviously, as kinetic energy is
progressively transformed into magnetic energy, the winding stage cannot
continue indefinitely and is expected to terminate when the
magnetic-field energy is in rough equipartition with the kinetic energy
stored in the differential rotation, so that further amplification is
energetically disfavoured. All of this can be directly deduced from the
left panel of Fig.~\ref{fig:mag_obs}, which reports the evolution of the
total magnetic energy (black solid line) and of its poloidal (red dashed
line) and toroidal (blue dotted line) components. Note that the KHI takes
place at merger and leads to an exponential growth of the toroidal and
poloidal components for $t-t_{\rm mer} \lesssim 5\,{\rm ms}$. After the
KHI is quenched and the merger has taken place, the poloidal component
essentially stops growing, while the toroidal component -- which was
absent before merger -- becomes the dominant component~\citep[see,
  \eg][for a detailed discussion of the magnetic-field evolution soon
  after the merger]{Kiuchi2015, Palenzuela_2022PRD, Chabanov2022}. It is
worth noting that the initial magnetic-field strength in our simulations
is considerably stronger than what is expected for old magnetized neutron
stars in a merger process. This has been shown to impact the topology of
the field after the initial amplification phase is
terminated~\citep{Aguilera-Miret2024}.

While these dynamics largely concern the magnetic field inside the
high-density region of the HMNS, the accretion-disk dynamics and dynamo
inside the disk will largely be governed by the MRI and shear-current
effects \citep{Christie2019b, Liska2018b, Jacquemin-Ide2023}. It is this
dynamics that leads to the growth of the poloidal component and to a
subsequent breakout as the one we will describe below.

\begin{figure*}
  \center
  \includegraphics[width=0.8\textwidth]{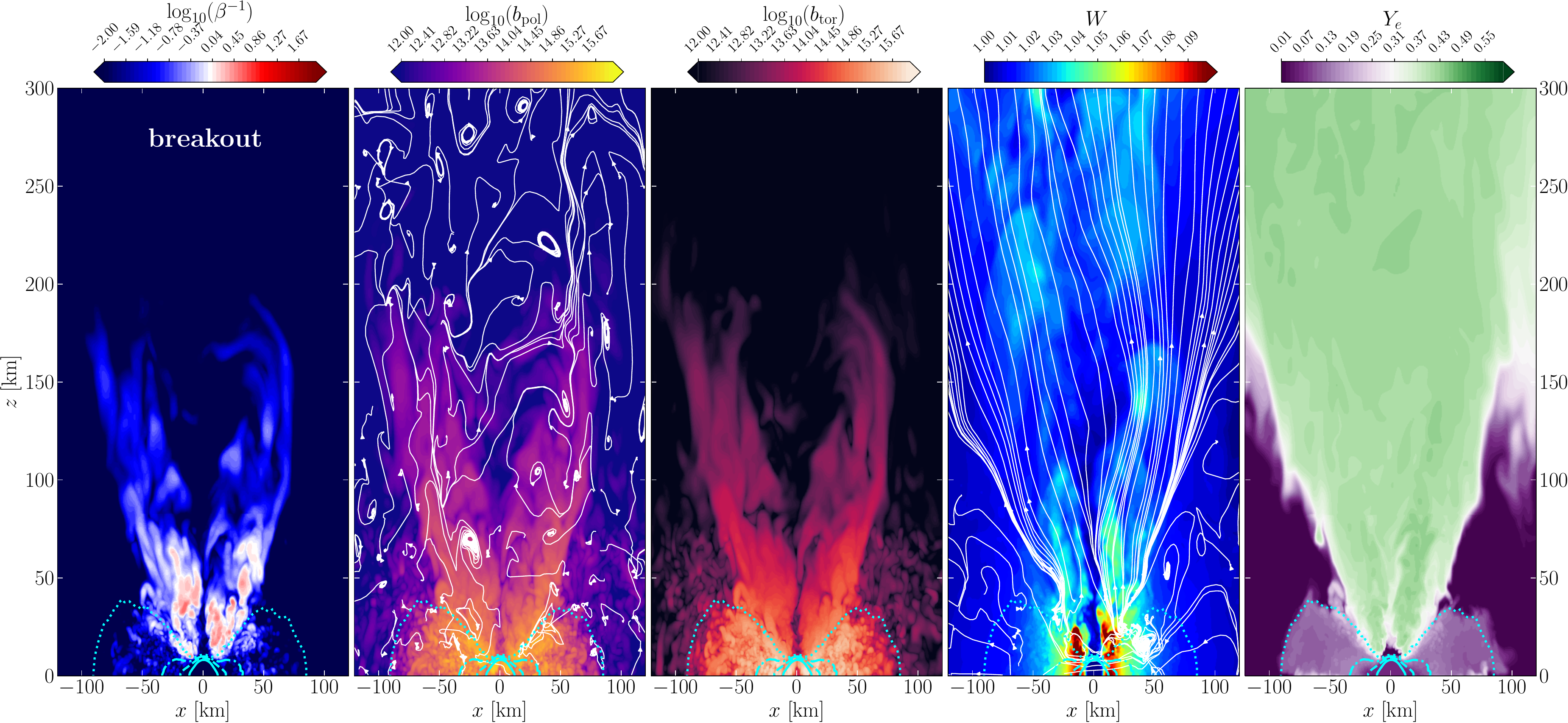}
  \vskip 0.25cm
  \includegraphics[width=0.8\textwidth]{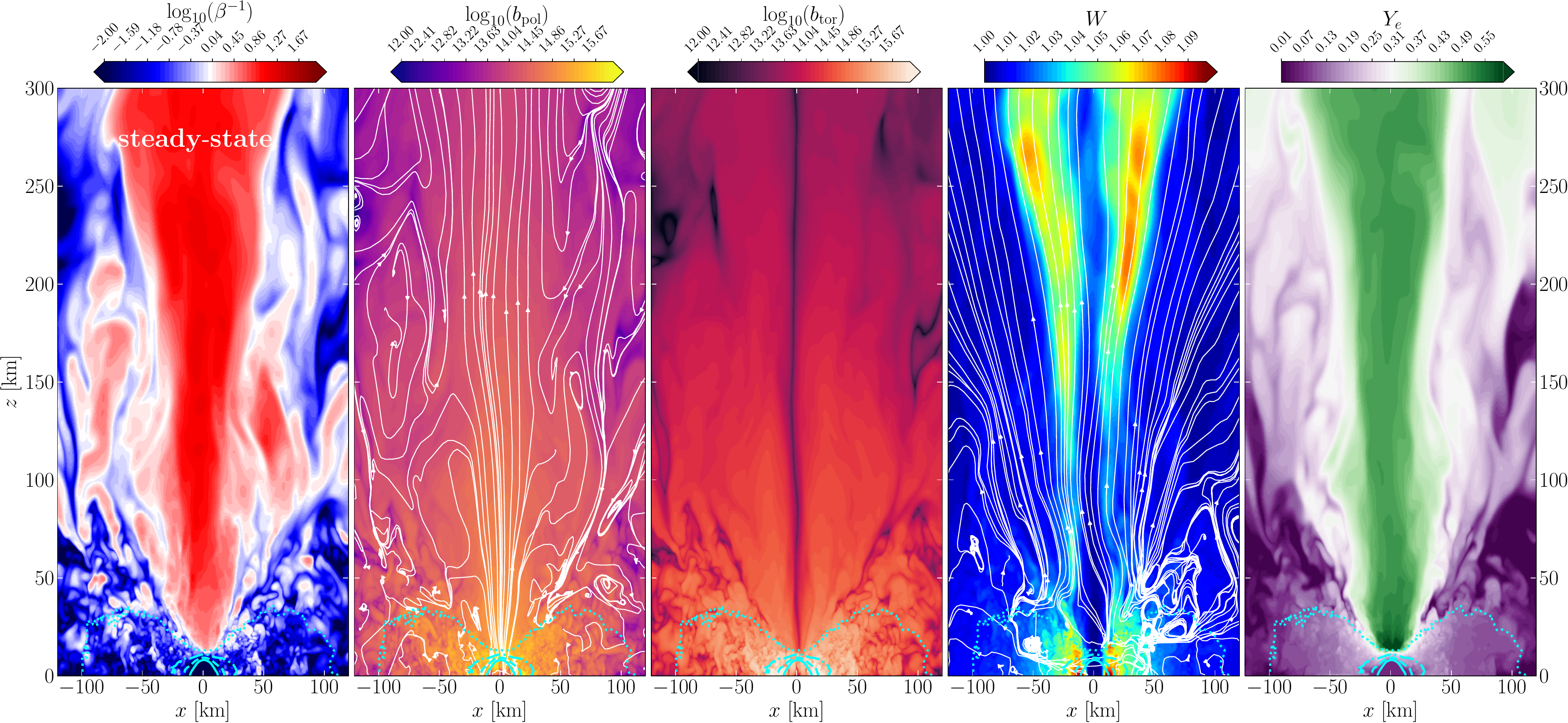}
  \caption{Two-dimensional distributions in the $(x,z)$ plane of various
    quantities computed in the simulations. Starting from the left, we
    report: the inverse plasma-$\beta$, \ie the ratio of the magnetic to
    thermal pressures the poloidal, $b_{\rm pol}$, and toroidal, $b_{\rm
      tor}$, components of the magnetic field in the comoving frame, the
    Lorentz factor, $W$, and the electron fraction, $Y_e$. While the top
    row refers to a time around the breakout, \ie at $t-t_{\rm mer} =
    39.8\,{\rm ms}$, the bottom row shows the evolution in the
    steady-state, \ie at $t-t_{\rm mer}=77.9\,{\rm ms}$. Light-blue
    contours refer to different rest-mass densities (dotted, dot-dashed,
    dashed and solid lines for $10^{10},\, 10^{12},\, 10^{13}$ and
    $10^{14}\,{\rm g/cm^3}$, respectively).}
  \label{fig:pans_break}
\end{figure*}

While the short-term dynamics described so far is well-known and has been
reported in a number of works, the subsequent long-term evolution is far
less studied in fully consistent simulations that account not only for
magnetic fields but also the influence of neutrino transport \citep[see,
  \eg][]{Combi2023}. To this scope, we report in the right panel of
Fig.~\ref{fig:mag_obs} the evolution of the EM luminosity as recorded via
a Poynting flux on a spherical 2-sphere with coordinate radius $\sim
1033\,{\rm km}$. What is clear from this panel is that the EM emission
from the remnant increases steadily and exponentially from the merger,
reaching a luminosity of $L_{_{\rm EM}} \simeq 10^{46}\,{\rm erg/s}$ at
about $15\,{\rm ms}$, when a sudden change takes place. We mark this
stage as the \textit{``breakout''} and will further discuss it in detail
below. After the breakout and for about $20\,{\rm ms}$, the EM luminosity
remains approximately constant to then increase by more than four orders
of magnitude to $L_{_{\rm EM}} \simeq 10^{50-51}\,{\rm erg/s}$ at $\sim
52\,{\rm ms}$; we mark this stage as the \textit{``eruption''}, which
should therefore be considered as the manifestation of the breakout for a
distant observer. This behavior is consistent with recent observations of
dynamo-driven breakout from HMNS \citep{Most2023, Most2023b,
  Kiuchi2023}. The subsequent evolution sees again an almost constant EM
emission up to the end of the simulation at $\sim 100\,{\rm ms}$; we will
refer to this as the ``steady-state'' stage. Under the conditions reached
at this state, the EM luminosity follows a simple scaling law in terms of
the initial magnetic field $B_0$, of the average equatorial radius of the
HMNS $R_e$ and of its peak rotation period $P$~\citep[see,
  \eg][]{Siegel2014}
\begin{equation}
  L_{_\mathrm{EM}} \! \simeq \!  10^{51} 
  \left(\frac{B_0}{10^{16}\,\mathrm{G}}\right)^{\!\!2} \!\!
  \left(\frac{R_e}{10^{6}\,\mathrm{cm}}\right)^{\!\!3} \!\!
  \left(\frac{P}{10^{-3}\,\mathrm{s}}\right)^{\!\!-1} \!\!\!\!
  \mathrm{erg/s}\, .
\label{eq:scaling}
\end{equation}
This behaviour is rather robust upon variations of the EOS and of the
topology of the magnetic field~\citep{Siegel2014}. Finally, before
closing this section, we should remark that what discussed so far in
terms of magnetic-energy evolution and EM luminosity applies
qualitatively also in the case in which neutrinos are completely ignored.
This is in part because neutrinos do not change the internal remnant
structure, such as the convectively stable stratification
\citep{Radice2023}, albeit they affect the degree of baryon loading of
the outflows \citep{Dessart2009}. The direct comparison with a
simulation not including neutrino transport is shown with gray lines in
Fig.~\ref{fig:mag_obs}, highlighting an important result of our
simulations: \textit{neutrinos do play a role in promoting and
  speeding-up the magnetic breakout, but are not necessary}. We will
return to this point below, when discussing more in detail the role of
neutrinos.

\subsubsection{Magnetically-driven winds}

To understand the physical processes responsible for the breakout and
eruption stages discussed above, it is useful to make use of the various
panels shown in Fig.~\ref{fig:pans_break}. Starting from the left, and
for both rows, we report: the ratio of the magnetic-to-thermal pressures
(this is also known as the inverse plasma-$\beta$), the poloidal ($b_{\rm
  pol}$) and toroidal ($b_{\rm tor}$) components of the comoving magnetic
field, the Lorentz factor ($W$)\footnote{It is possible to define an
asymptotic Lorentz factor $W_{\infty}:=-(h+b^2/\rho)u_t$, where $h$ is
specific enthalpy and $b$ the magnetic-field strength, and $u_t$ the
covariant time component of the fluid four-velocity. In our simulations
we have found that $W_{\infty}/W \lesssim 2$ on average}, and the
distribution of the electron fraction ($Y_e$). Importantly, the top row
refers to a time around breakout, \ie at $t-t_{\rm mer}=39.8\,{\rm ms}$,
while the bottom row shows the evolution in the steady-state, \ie at
$t-t_{\rm mer}=77.9\,{\rm ms}$.

When contrasting the top and bottom panels of Fig.~\ref{fig:pans_break},
it is clear that something dramatic takes place at the breakout and that
this is most evident in the changes that occur in the pressure ratio
[indeed, $\log_{10}(\beta^{-1}) \ll -2$ everywhere before breakout; see
  Appendix~\ref{sec:app_A}] and in the topology of the magnetic
field. Equally clear is that after breakout the magnetic pressure $p_{\rm
  mag}$ dominates over the thermal pressure $p$ in the whole polar
region, underlining the magnetically-dominated nature of the outflow and
its large electron-fraction content (rightmost panel). Similarly, the
open poloidal magnetic-field lines, which were about to be formed and
were anchored in the disk at breakout, subsequently fill the polar region
and are anchored in the differentially rotating HMNS (the total magnetic
field lines are also anchored in the disk). The magnetically-driven wind
that ensues exhibits a clear jet-like structure, with velocities that
are high close to the surface of the HMNS and become increasingly larger
as the outflow moves to larger distance.

Furthermore, the wind has a characteristic angular structure where the
flow is slower closer to the rotation axis, where the toroidal field is
necessarily less strong due to the reduced winding. This feature has been
observed in numerous other works describing similar phenomenology, across
different systems and mass scales~\citep[see,
  \eg][]{Combi2023,Most2023,Bamber2024}. Finally, while the magnetically
driven wind has a clear collimated structure, the corresponding outflow
is only \textit{mildly relativistic}, with Lorentz factors that are
everywhere $W\lesssim 1.2$. We caution that the value we find is lower
than that recently reported by \citet{Kiuchi2023}, \ie $W\simeq
10-20$. This may indicate that the disk breakout scenario presented does
not yet produce strong enough fields compared to simulations that fully
capture the MRI also inside the HMNS or that longer evolution timescales
are needed to reach larger Lorentz factors. When this highly magnetized
and energetic wind reaches the detector at $t-t_{\rm mer} \sim 50\,{\rm
  ms}$, it will manifests itself in a sudden jump in the EM luminosity,
exactly as reported in the right panel of Fig.~\ref{fig:mag_obs}. As we
will comment below, the outbreak of this magnetically-driven wind also
sweeps material in the funnel region, increasing the neutrino luminosity.

\begin{figure*}
  \center
  \includegraphics[width=0.45\textwidth]{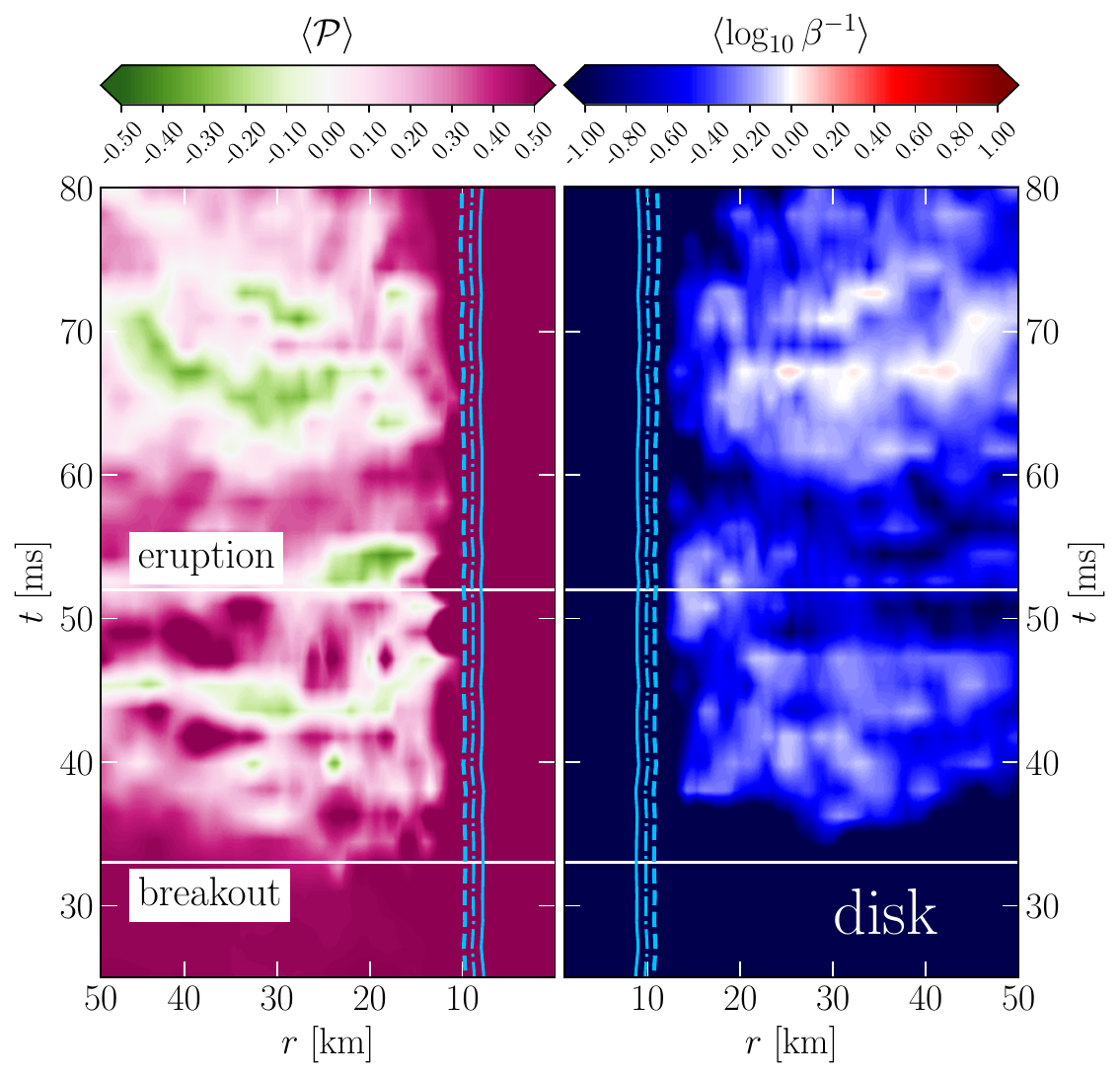}
  \hskip 0.5cm
  \includegraphics[width=0.45\textwidth]{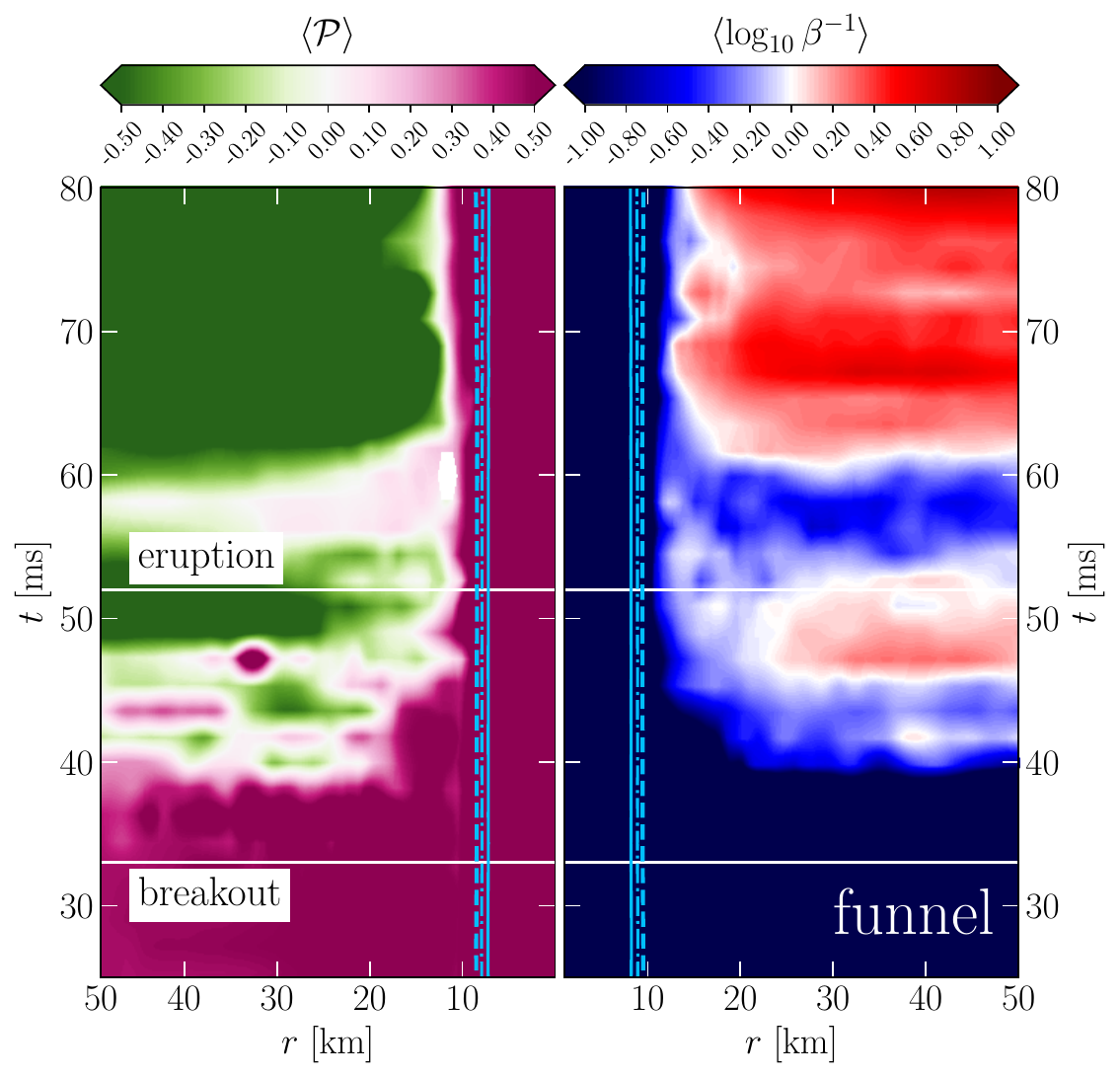}
  \caption{\textit{Left panel:} Spacetime diagram of the averaged inverse
    plasma-$\beta$ (right part) and of the averaged Parker-stability
    criterion (left part); both parts report data associated with the
    ``disk''. \textit{Right panel:} The same as in the left panel but for
    the region associated with the ``funnel''. Both panels show the time
    of ``breakout'' marked with a horizontal white line and blue contours
    for different rest-mass densities (dashed, dot-dashed and solid lines
    for $10^{14},\, 10^{13.5}$ and $10^{13}\,{\rm g/cm^3}$; the latter
    can be taken as to mark the surface of the HMNS).}
  \label{fig:spacetime_torus}
\end{figure*}

\subsection{$\alpha$-$\Omega$~dynamo, Parker criterion, and breakout}
\label{sec:parker}

If the phenomenology discussed so far has a simple and consequential
interpretation where, at one point in the evolution, a magnetic breakout
takes place yielding a magnetically driven flow that fills the polar
region and enhances the EM emission, the actual origin of the instability
leading to the breakout still requires a proper explanation. We have
mentioned above that after the turbulent amplification of the magnetic
field via the KHI, the shearing flows in the HMNS transfer the kinetic
and binding energy from the fluid over to the magnetic fields via two
main mechanisms. The first one is the simple magnetic winding via the
differential rotation in the HMNS leading to the growth of a
predominantly toroidal magnetic field. The second one, instead, involves
the MRI and produces mostly poloidal magnetic field (see left panel of
Fig.~\ref{fig:mag_obs}). Recent simulations of BNS mergers at ultra-high
resolution~\citep{Kiuchi2023} have shown the presence of an MRI-driven
$\alpha$-$\Omega$~dynamo~\citep{Bonanno:2003uw} in the remnant HMNS,
which can lead to the growth of the magnetic field close to the HMNS
polar cap and ultimately to a breakout of the field, although the
location of the breakout (disk, star or both) will depend crucially on
the dynamo processes \citep{Most2023, Most2023b}.

Our resolution of $\sim 200~{\rm m}$ is high but does not allow us to
resolve the wavelength of the fastest-growing MRI mode within the
high-density region of the HMNS~\citep{Siegel2013}, and indeed we do not
observe any growth of the magnetic field in regions of rest-mass density
$\rho \gtrsim 10^{12}~{\rm g/cm^3}$ besides linear winding. However, less
severe resolutions requirements are needed to resolve the MRI in the
inner region of the accretion disc, where the density is $\gtrsim 10^9 \,
{\rm g/cm^3}$, which in fact we do resolve accurately. This allows us to
point out that a mechanism similar to that reported
by~\citet{Most2023,Combi2023,Kiuchi2023} can act self-consistently
\textit{also} in the disk and ultimately lead to a buoyant instability
with consequent breakout of the magnetic field.

To demonstrate this mechanism, and disentangle the various parts of the
flow involved in this complex process, we distinguish the contributions
coming from two distinct regions: (i) the \textit{``funnel''}, \ie a
region near the equatorial plane with radial extent $1< r/{\rm km} < 50$
and polar excursion $0 < \theta < \pi/6$; (ii) the \textit{``disk''}, \ie
a region with the same radial extent but around the polar cap, \ie with
$1 < r/{\rm km} < 50$ and $\pi/6 < \theta < \pi/2$. We then use
Fig.~\ref{fig:spacetime_torus} to report the spacetime diagrams for two
quantities relevant to the breakout dynamics: the inverse plasma-$\beta$,
and the ``Parker-stability criterion'' $\mathcal{P}$, namely, a measure
of the onset of the magneto-convective instability~\citep{Parker1966}
\begin{equation}\label{eq:parker_criterion}
  \mathcal{P} := \frac{d\log{p}}{d\log{\rho}} - 1 -
  \frac{\beta^{-1}\left( 1 + 2\beta^{-1} \right)}{2 + 3 \beta^{-1} }\,.
\end{equation}
Both quantities are displayed on the $(x,z)$ plane as an average over the
azimuthal angle and refer to either the disk (left panel) or to the
funnel (right panel). Within each panel, we distinguish the evolution of
the Parker criterion (left side) from that of the inverse plasma-$\beta$
(right side). Furthermore, we mark with light blue lines the worldlines
of fluid elements at reference rest-mass densities, \ie
$\log_{10}(\rho)=(13,14,15)\,{\rm g/cm^3}$ (solid, dashed, and dot-dashed
lines); in this way, the thick solid line can be taken to mark
approximately the surface of the remnant HMNS~\citep[see][for a detailed
  discussion]{Cassing2024}.

The four spacetime diagrams in Fig.~\ref{fig:spacetime_torus} clearly
help realizing that the breakout time marks the moment when the Parker
criterion changes sign and hence a buoyant instability is allowed to take
place in the disk. The two panels in Fig.~\ref{fig:spacetime_torus} also
highlight that the conditions in the breakout first take place in the
disk and only a few milliseconds later in the funnel. Also apparent is
the fact that while the outer regions of the disk are no longer
matter-dominated after breakout (dark blue turns to white), it is really
the funnel that experiences the largest excursion from a matter-dominated
to a magnetically-dominated regime (dark blue turns to dark red); a
similar phenomenology has been reported also in other simulations under
different approximations but comparable conditions~\citep{Kiuchi2023,
  Most2023,Most2023b}.

\begin{figure*}[t]
  \center
  \includegraphics[width=0.95\textwidth]{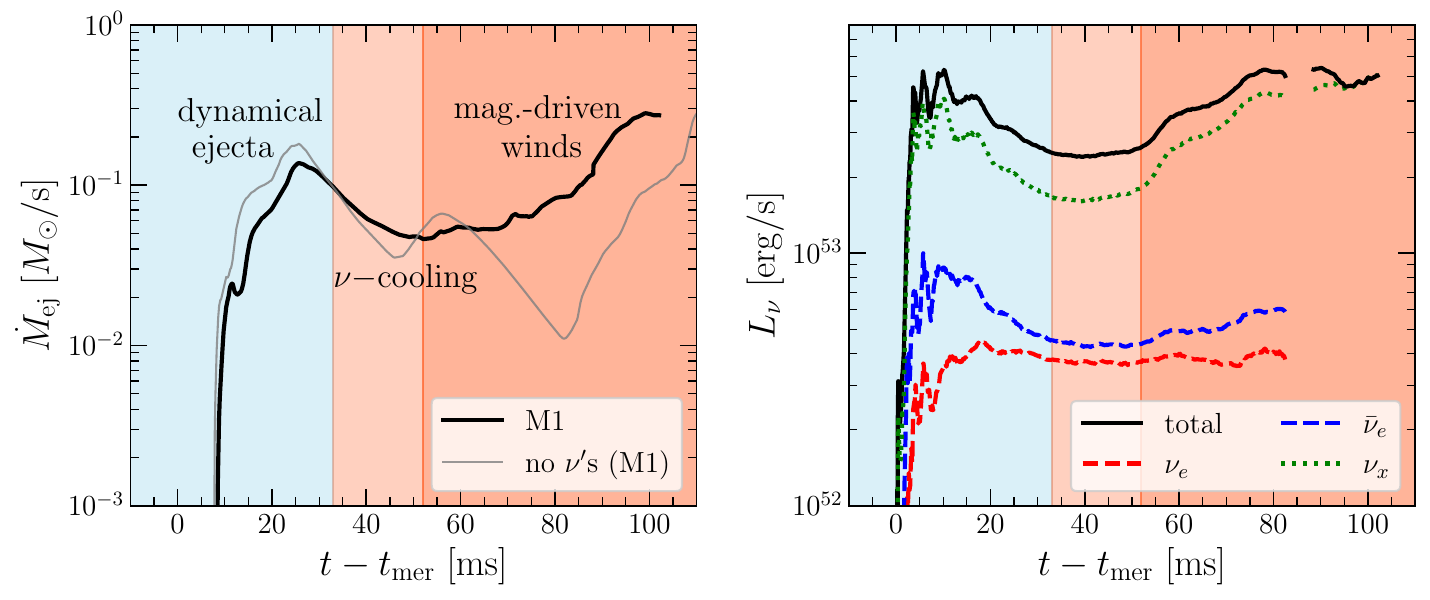}
  \caption{\textit{Left panel:} Evolution of the ejection rest-mass rate
    ${\dot M}_{\rm ej}$, showing first the peak of the dynamical ejecta
    and subsequently that the cooling by neutrinos reduces the efficiency
    of ${\dot M}_{\rm ej}$ (the color-shading is the same as in
    Fig.~\ref{fig:mag_obs}). After eruption, however, the mass-ejection
    rate increases thanks to the launching of the magnetically-driven
    winds; a gray line shows the corresponding evolution when no
    neutrinos are taken into account (no ${\rm M1}$). \textit{Right
      panel:} Evolution of the total neutrino luminosity (black solid
    line) together with the contributions from heavy neutrinos (green
    dashed line), electron neutrinos and antineutrinos (red and blue
    dashed lines, respectively).}
  \label{fig:other_obs}
\end{figure*}

\subsection{Mass ejection and the role of neutrinos}
\label{sec:role_neutrinos}

The magnetic-field breakout and eruption also has an impact on the amount
of mass that is ejected from the merger remnant. This is illustrated in
the left panel of Fig.~\ref{fig:other_obs}, where we report the rest-mass
ejection rate computed as the surface integral of the unbound mass flux
(according to the standard ``Bernoulli'' criterion $h u_t <
-1$)\footnote{It is possible to add magnetic contribution via a term
proportional to $b^2\,u_t/\rho$ but this hardly changes the plot given
the small relative value of the magnetization.} at the same detector
where we measure the Poynting flux. The evolution prior to the eruption
shows the well-understood initial burst of unbound material that is
commonly referred to as the ``dynamical ejection'', \ie the launching of
high specific-entropy material accelerated by the collision of the two
merging stars and typically occurring in multiple waves as the stellar
cores bounce repeatedly~\citep{Bovard2017, Bernuzzi2020, Dietrich2016,
  Neuweiler2022, Papenfort2018, Most2020e, Zappa2023}. This initial phase
of dynamical ejection is followed by a sustained wind of unbound matter
(see Fig.~\ref{fig:other_obs}), which is largely caused by neutrino
energy deposition in the funnel (see also the discussion in
Appendix~\ref{sec:app_B}). This ``neutrino-driven'' wind
\citep{Dessart2009}, which kicks-in shortly after merger, dominates at
intermediate times between the end of the dynamical ejection phase and
the onset of the magnetic ejection and, as we will discuss next, plays an
important role in clearing the funnel and promoting the proper conditions
for the launching of the outflow.

As can be observed from Fig.~\ref{fig:mag_obs}, after breakout the
mass-ejection rate slowly grows together with the Poynting flux,
indicating that the unbound material becomes progressively more
magnetized. Once the magnetic field completely fills the polar region
above the HMNS and the funnel becomes magnetically dominated, the
Poynting flux luminosity increases steeply and the wind becomes
essentially magnetically driven, forming a jet-like
configuration~\citep{Lynden-Bell1996, Rezzolla:2011, Shibata2011b}, which
can clearly be observed in Fig.~\ref{fig:pans_break}. This phenomenology
is similar to what is described by~\citet{Most2023b} in the case of
maximum amplification of the turbulent magnetization, where a steady,
high-Poynting flux outflow is observed. However, we can highlight for the
the first time that a Blandford-Payne magneto-centrifugal
mechanism~\citep{Blandford:1982di} is also active and contributing to the
the acceleration and collimation of the outflow and quantifiable in terms
of the magnetic lever-arm contribution to the fluid energy [see the
  detailed discussion of Eq.~\eqref{eq:Bern+} and
  Figs.~\ref{fig:bernoulli_nu}-\ref{fig:bernoulli_nonu} in
  Appendix~\ref{sec:app_B}].

Overall, the values measured in our simulations are $\dot{M}_{\rm ej}
\sim 4-5 \times 10^{-2}\,M_{\odot}/{\rm s}$ and hence about a factor
two-three larger than those estimated by~\citet{Gill2019}. The
magnetically-driven wind has generally larger speeds than those of the
neutrino-driven wind, with the former having Lorentz factors $W\lesssim
1.05$ and terminal Lorentz factors $W_\infty := - h\,u_t \lesssim 1.2$,
while the latter reaches Lorentz factors of $W\sim 1.1$ and terminal
Lorentz factors of $W_\infty \sim 1.5$.

The merger obviously lead also to a dramatic change in the
thermodynamical state of the neutron stars, raising their temperature
from a few eV minutes prior to merger to tens of MeV in the course of a
few milliseconds after the merger. As the HMNS reaches a quasi-stationary
MHD equilibrium, so does its temperature stratification, which is
represented by a dense and relatively cold core surrounded by a hotter
ring of rapidly rotating matter, whose temperature gradually decreases
when moving outwards to smaller rest-mass densities~\citep[see,
  \eg][]{Kastaun2014, Hanauske2016}. As a result of this quasi-stationary
thermodynamical equilibrium, a copious amount of neutrinos are produced
and emitted.

In turn, neutrinos are also responsible for the ejection of substantial
amounts of matter via neutrino-driven winds, which have been the subject
of numerous studies in the context of BNS mergers~\citep{Dessart2009,
  Perego2014, Martin2015, Fujibayashi2017, Sekiguchi2017}. At the risk of
oversimplification, these neutrino-driven winds consist in polar outflows
with relatively high electron fraction ($0.3 \leq Y_e \leq 0.5$) and
mildly relativistic speeds ($W \lesssim 1.1$). They are caused by
transfer of momentum from the neutrinos to the matter in the outer layers
of the star and of the high-density regions of the disk. As a result,
they contribute to the large mass-ejection rates measured in our
simulations prior to breakout. Indeed, until breakout, neutrinos can be
considered as the main ``driver'' of the mass outflow. This can be
readily deduced from the rightmost panel in the top row of
Fig.~\ref{fig:pans_break}, which shows the velocity field as a
stream-plot. Note the presence of an ordered outflow in the funnel where
matter is unbound (see discussion in Appendix~\ref{sec:app_B}), where
however the magnetic field is still very low (see leftmost panel of
Fig.~\ref{fig:pans_break}) and the thermal pressure dominates over the
magnetic pressure ($\beta^{-1} \ll 1$). Under these conditions, the source
of acceleration of the unbound material in the funnel is not magnetic in
origin and hence the outflow is neutrino driven.

The evolution of the neutrino luminosities is shown in the right panel of
Fig.~\ref{fig:other_obs}, where we report the total luminosity (black
solid line) and also the contributions coming from heavy neutrinos (green
dashed line), electron neutrinos and antineutrinos (red and blue dashed
lines, respectively). Note how the main component of the total neutrino
luminosity, that reaches values $L_{\nu} \sim 4 \times 10^{53}\,{\rm
  erg/s}$, comes from the heavy neutrinos, whose luminosity is more than
twice that of electron neutrinos and antineutrinos\footnote{We note that
some of the M1 data in the final part of the simulation was unfortunately
corrupted, likely due to a lack of numerical dissipation of the M1 scheme
in the optically thin limit. This explains the lack of data in the final
part of the evolution in the right panel of Fig.~\ref{fig:other_obs},
especially for electron neutrinos and antineutrinos.}. Of course, these
considerable losses in energy come at the expense of the thermal and
kinetic energies of the system. The former manifests itself in a cooling
of the remnant and a corresponding gradual reduction of the neutrino
luminosity at about $10\,{\rm ms}$ after merger. Moreover, the energy
carried away by neutrinos leads to a reduced dynamical ejection of
matter. Indeed, including neutrino transport in simulations of BNS
mergers generally reduces the total amount of dynamically ejected
material by up to a factor of two compared to simulations that exclude
neutrino effects, as can be evinced by comparing the gray and black
curves in the left panel of Fig.~\ref{fig:other_obs}~\citep[see
  also][]{Zappa2023}. In addition, the absence of neutrino essentially
quenches the mass-ejection rate after the dynamical-ejection episode at
$\sim 15\,{\rm ms}$ after the merger, and will be necessary to wait for
the subsequent breakout, occurring much later and at about $100\,{\rm
  ms}$ after the merger, to see a significant increase the mass-ejection
rate.

Before breakout, neutrinos also play the very important role of
``cleaning up'' the funnel, namely, of reducing the baryon loading in the
funnel and hence promote the breakout of magnetic field from the torus
surface. This is clearly correlated with the evolution in the neutrino
luminosity, that stops decreasing between breakout and eruption and then
increases after eruption (see the black solid line in the right panel of
Fig.~\ref{fig:other_obs}). This is simply the consequence of the fact
that the funnel has now been cleaned up of baryons (hence reducing
neutrino capture and scattering) and neutrinos can be emitted more
easily.

When neutrinos are ignored and the cleaning up does not take place, the
polar regions of the remnant are filled with baryon-rich material that
cannot be blown out via a neutrino-driven wind. As a result, the mass
stops being ejected after the initial dynamical ejection and, as it can
be appreciated from the gray line in the left panel of
Fig.~\ref{fig:other_obs}, the mass-ejection rate steadily decreases till
$t-t_{\rm mer} \simeq 85\,{\rm ms}$, when it is revived by the magnetic
breakout. More importantly, the baryon pollution outside the HMNS
prevents the expulsion of magnetic field from the disk surface, which
takes place at $t-t_{\rm mer} \simeq 35\,{\rm ms}$, when neutrinos are
accounted for. Much larger magnetic fields will have to be produced via
winding to reach the critical Parker-stability limit and start launching
and magnetically driven outflow. This is particularly clear when looking
at the right panel of Fig.~\ref{fig:mag_obs}, where the solid gray line
shows how the EM luminosity is essentially constant at the eruption of
magnetic field takes place at $t-t_{\rm mer} \simeq 80\,{\rm ms}$ (the
breakout in this case happens somewhat earlier, \ie at $t-t_{\rm mer}
\sim 50\,{\rm ms}$; see also Fig.~\ref{fig:spacetime_torus_nom1} in
Appendix~\ref{sec:app_B}). This result, should help resolve the on-going
debate within the literature and concerned on whether neutrinos or
magnetic fields are responsible for the generation a powerful and
collimated but non-relativistic outflow~\citep{Ciolfi2020_a,
  Moesta2020}. The black and gray lines in the right panel in
Fig.~\ref{fig:mag_obs} address this question beyond debate and clearly
show what we have already anticipated: neutrinos play an important and
promoting role in launching a magnetically-driven outflow but are not
strictly necessary. To the best of our knowledge this is the first time
that point is addressed with self-consistent GRMHD and M1
neutrino-transfer calculations \citep[see also][for one-moment
  approaches]{Combi2023}.

\section{Conclusion}
\label{sec:conclusions}

Whether the metastable remnant of a BNS merger can launch a relativistic,
magnetized and collimated outflow has been a matter of debate over the
last few years, with different approaches reaching different
conclusions. Examples include the development of magnetically-driven
outflows for simplified microphysics treatments
\citep[\eg][]{Ciolfi2020_a, Pavan2023, Bamber2024}, or simplified
magnetic-field evolutions with more realistic neutrino microphysics
\citep[\eg][]{Moesta2020, deHaas2022, Curtis2023}), or both
\citep{Combi2023}. Recently, several works have also investigated the
importance of dynamo processes \citep{Most2023, Most2023b, Kiuchi2024} as
a prerequisite for field breakout and the formation of a collimated
outflow. With the goal of clarifying this picture in particular with
regards to the importance and role of neutrinos, we have carried out
long-term simulations of the inspiral, merger and post-merger evolution
of a BNS system in which both magnetic fields and neutrino absorption and
emission are properly described. In addition, we have replicated the same
simulation setup neglecting neutrinos, so as to highlight and distinguish
the role played by magnetic fields and neutrinos.

This analysis has allowed us to reach a number of conclusions, which we
briefly summarise below. First, right after merger an intense neutrino
emission takes place that generates a neutrino-driven wind and
``clears-up'' the polar region above the remnant from matter. Second,
after $\sim~50\,{\rm ms}$ from the merger, a dynamo driven by the MRI
develops in the densest regions of the disk and leads to the breakout of
the magnetic field lines in the polar region. Third, the breakout, which
can be directly associated with the violation of the Parker-instability
criterion and is responsible for a collimated, magnetically driven,
Poynting-flux dominated outflow with only mildly relativistic velocities.
Based on higher Lorentz factors reached in very high resolution studies
\citep{Kiuchi2024}, we speculate that the details may strongly depend on
the regions where the dynamo is active and breakout eventually happens.
The outflow we observe is partly accelerated and collimated via a
Blandford-Payne mechanism driven by the open field lines anchored in the
inner-disk regions. Finally, contrasting the simulations with and without
neutrino transport we were able to ascertain that the physical mechanism
of magnetic-field breakout is robust and takes place even without
neutrinos. The latter, however, lead to a cooler, geometrically thinner
disk and a lower degree of baryon pollution in polar regions, hence
anticipating the occurrence of the breakout and yielding a larger
Poynting flux.

The overall conclusion of this work is that while magnetically-driven and
magnetically-confined outflows can and are produced robustly from the
metastable merger remnant of BNS mergers, the associated jet-like
outflows launched via disk-driven breakout are only mildly relativistic
nature. In the absence of additional field amplification in denser
regions \citep{Kiuchi2024} (however, see also
\citet{Aguilera-Miret2024}), the mechanism presented here might be
insufficient to account for outflows responsible for powering short
gamma-ray bursts. However, a more conclusive statement will have to wait
for new simulations of this type at even larger resolutions, or using
suitably tuned effective mean-field dynamo models
\citep{Most2023b}. These are needed in order to properly resolve the MRI
also in the regions of the HMNS with the highest magnetic fields and
densities, which are precluded from our analysis with the present
resolution. While these regions are much smaller in size than the those
involved in the magnetic breakout from the disk (\ie $\sim 10\,{\rm km}$
vs $\sim 50\,{\rm km}$) and will not prevent the breakout observed here,
it is possible that they will also contribute to the breakout and hence
to the launching of collimated and magnetically-driven winds from the
remnant. We will assess this possibility in future studies.

\newpage  
\appendix  

\section{Simulations without neutrino transfer}
\label{sec:app_A}

As already mentioned in several instances in the main text, the study
carried out here and the results obtained are corroborated by a parallel
set of simulations in which neutrino radiative transfer via an M1
approach is neglected. Of course, we do not neglect neutrinos because
unimportant. Rather, since this is still a matter of a lively debate, we
do so in order to disentangle their role from that of magnetic fields and
assess whether they are necessary to launch a magnetically-driven and
collimated outflow or not.

\begin{figure*}[h!]
  \center
  \includegraphics[width=0.8\textwidth]{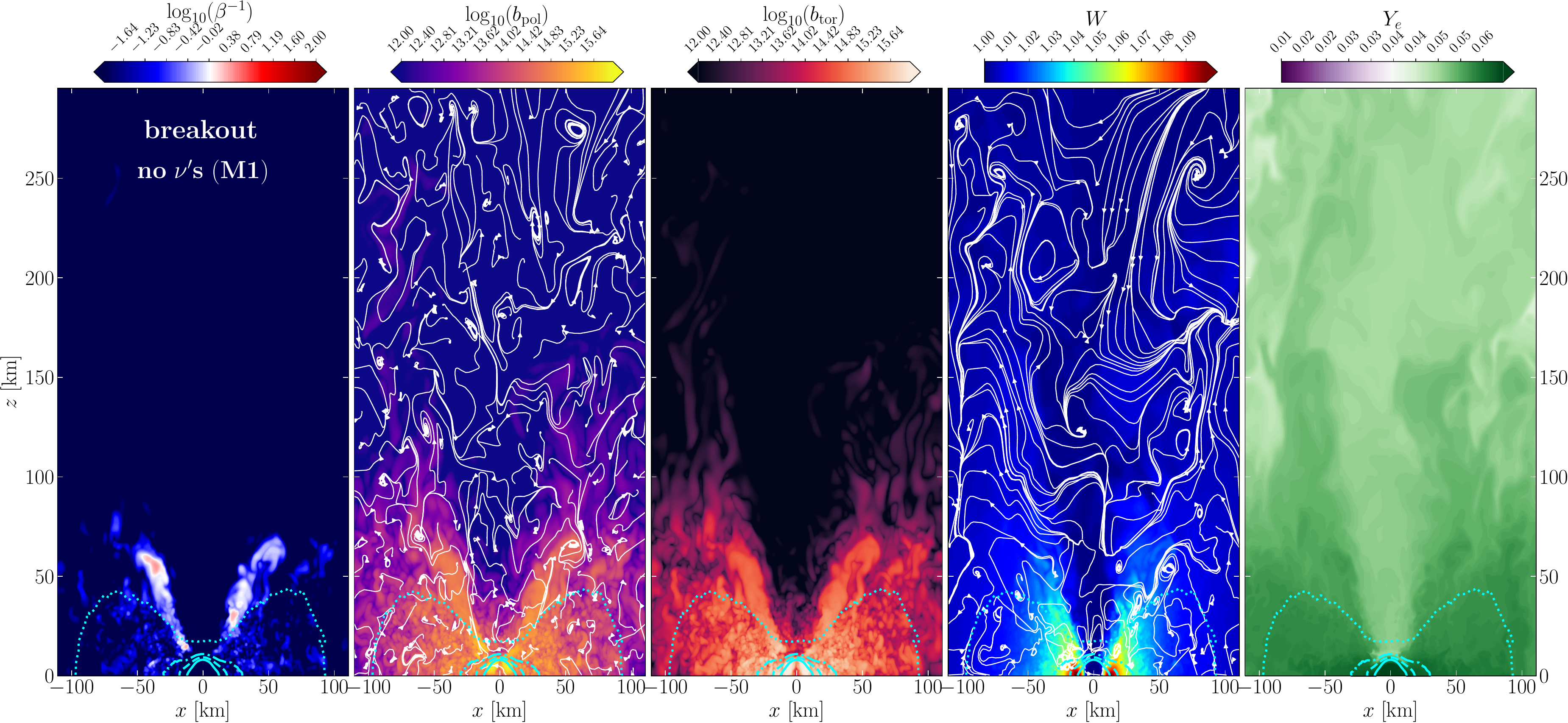}
  \vskip 0.25cm
  \includegraphics[width=0.8\textwidth]{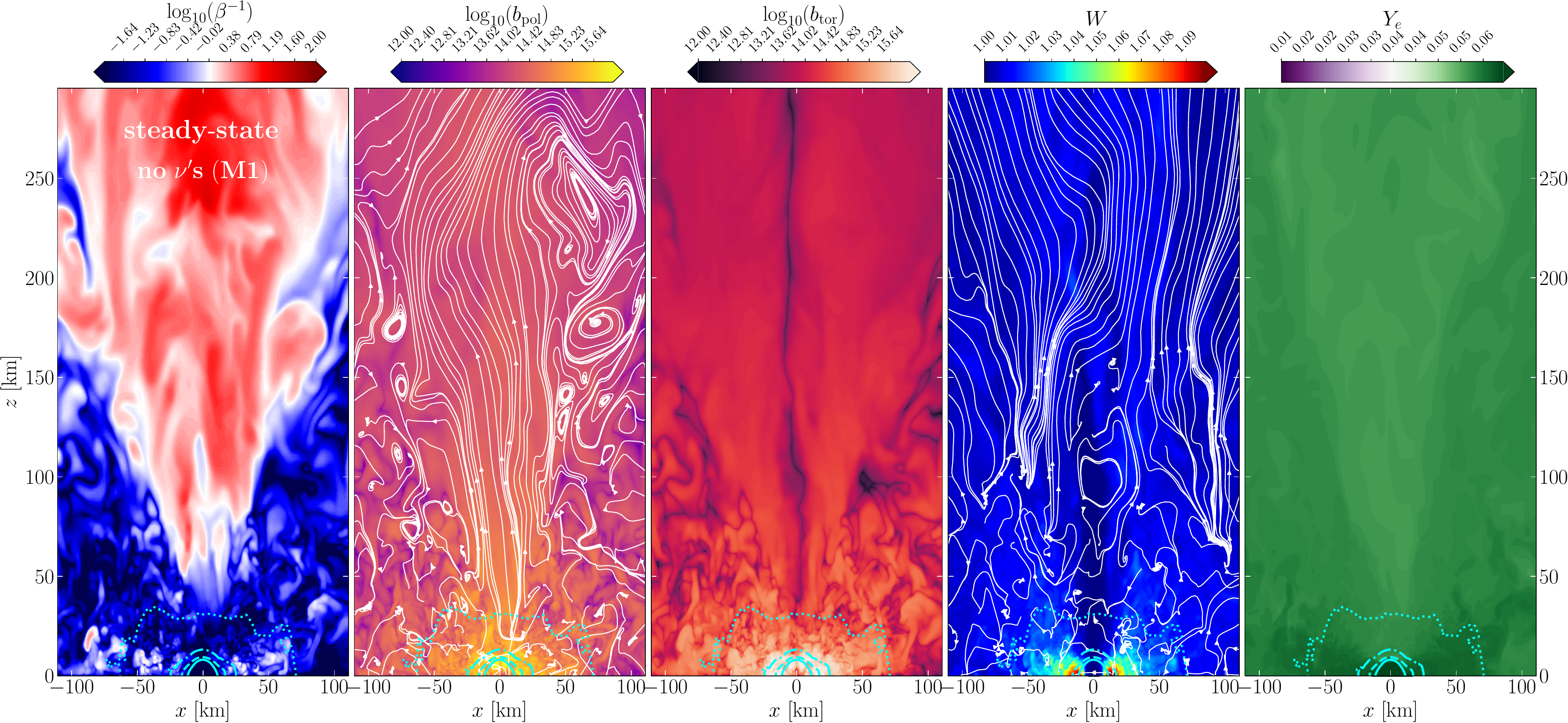}
  \caption{The same as in Fig.~\ref{fig:pans_break} but for a GRMHD
    simulation without neutrino transfer. Note that the breakout still
    takes place but at much later times ($t-t_{\rm mer} \simeq 50\,{\rm
      ms}$) and that the steady-state ($t-t_{\rm mer} \simeq 100\,{\rm
      ms}$) shows weaker, less collimated and slower outflows. Note the
    considerably different rage in which $Y_e$ is shown.}

  \label{fig:pans_break_nom1}
\end{figure*}

Figure~\ref{fig:mag_obs} reports with gray lines of various type the
evolution of the corresponding quantities when neutrinos are neglected
(no ${\rm M1}$) and clearly shows that neglecting neutrinos does not
introduce a qualitative change in the evolution of the magnetic energy
(left panel) or of the EM luminosity (right panel). At the same time, and
from a more quantitative point of view, it also highlights that neutrinos
do actually facilitate the breakout, anticipating it by several tens of
milliseconds.

Another qualitative and quantitative comparison between the two scenarios
is offered in Fig.~\ref{fig:pans_break_nom1}, which is the same as in
Fig.~\ref{fig:pans_break}, but for a GRMHD simulation without neutrino
transfer, and where it should be noted the considerably different range
in which $Y_e$ is shown. Note that the breakout still takes place but at
much later times ($t-t_{\rm mer}\simeq 50\,{\rm ms}$) and that before
breakout both the poloidal and toroidal magnetic fields are stronger than
when considering neutrinos. This is because the remnant is not losing
kinetic and binding energy via neutrinos and hence it is more efficient
in converting the latter two into magnetic energy, thus amplifying the
magnetic-field strength after merger (see also the left panel of
Fig.~\ref{fig:mag_obs}). In addition, the lack of neutrinos and the
corresponding neutrino-driven winds implies that when a steady-state is
reached ($t-t_{\rm mer}\simeq 100\,{\rm ms}$), a larger baryon pollution
is present in the remnant's polar regions and that the corresponding
poloidal magnetic-field component will be more irregular and with a
smaller large-scale component. As a result, the magnetically-driven
outflow will be shows weaker, less collimated and slower. 

\begin{figure*}[h!]
  \center
  \includegraphics[width=0.45\textwidth]{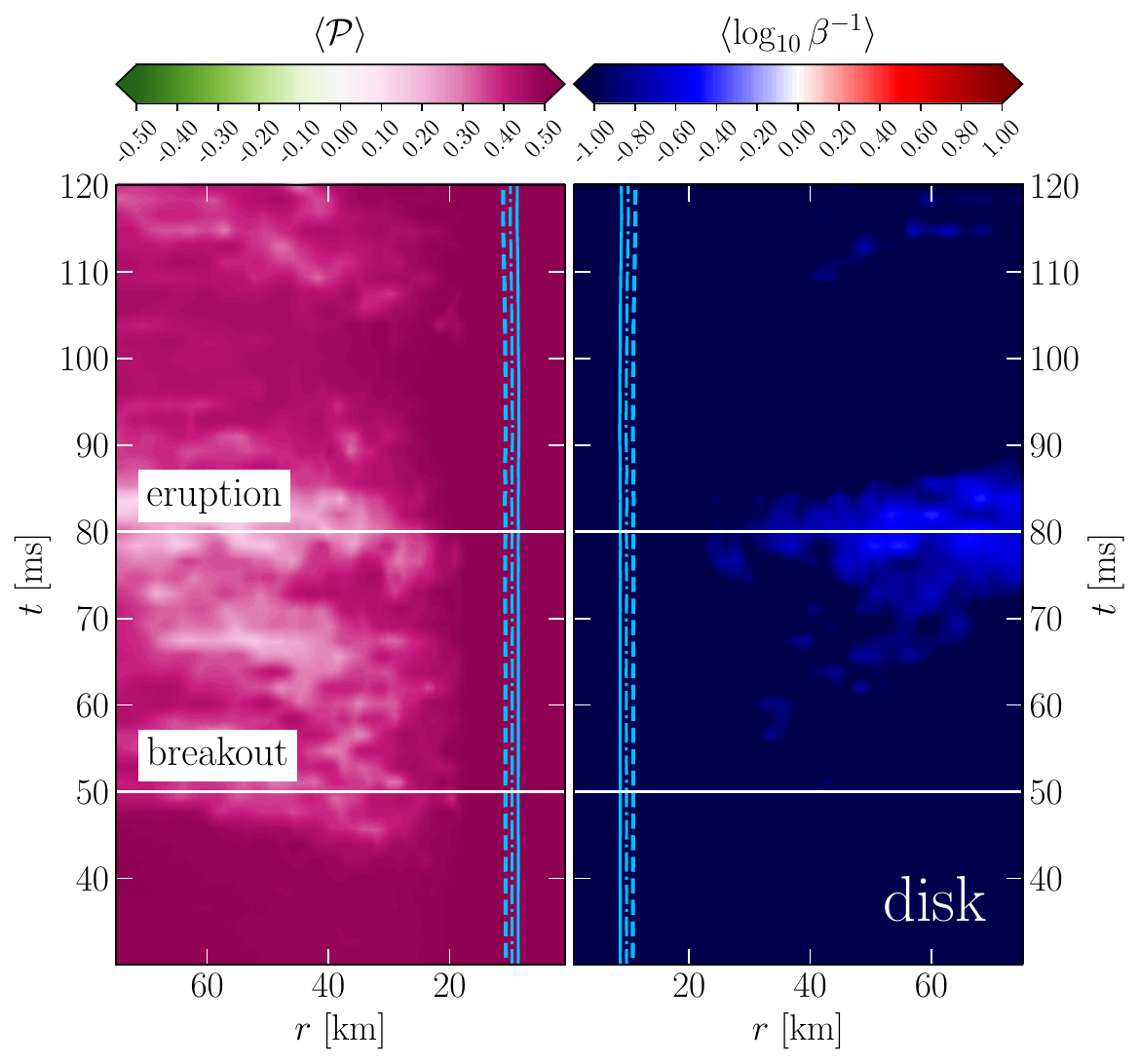}
  \includegraphics[width=0.45\textwidth]{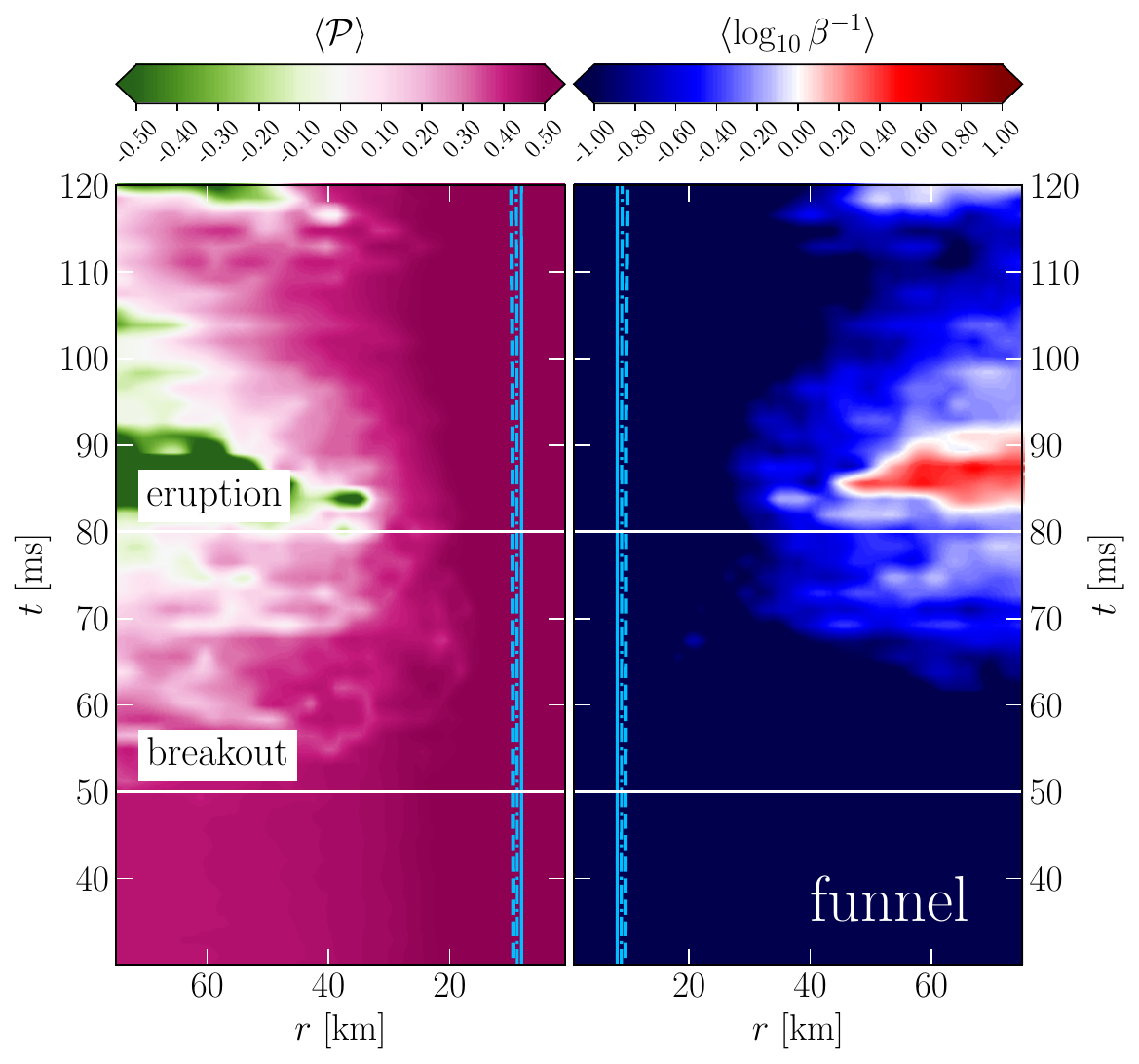}
  \caption{The same as in Fig.~\ref{fig:spacetime_torus} but for a GRMHD
    simulation without neutrino transfer.}
  \label{fig:spacetime_torus_nom1}
\end{figure*}

Figure~\ref{fig:spacetime_torus_nom1} provides another point of
comparison by offering the equivalent of the spacetime diagrams presented
in Fig.~\ref{fig:spacetime_torus_nom1} in the case when neutrino are
neglected. Note that the qualitative behaviour is the same and that a
sharp transition takes place when the Parker-stability criterion is
violated. Also in this case, the inverse plasma-$\beta$ goes from being
very small to being very large. However, the time when this takes place
is considerably later and the breakout can here be seen to start at
$t-t_{\rm mer} \simeq 50\,{\rm ms}$.

In summary, the discussion presented here, combined with that in the main
text, reinforces a point we have already highlighted repeatedly: while
neutrinos promoting and speed-up the magnetic breakout, they are not
necessary. When neglected, a magnetically-driven wind will be produced
nevertheless, although it will be less collimated and with smaller
Lorentz factors. In either case, it is difficult to associate these
outflows with the ultra-relativistic jets expects in short gamma-ray
bursts.

\section{Decomposition of outflow energetics}
\label{sec:app_B}

Having in mind the description of the mass-ejection dynamics discussed in
the main text, we now switch to a more quantitative discussion of the
energetics of the magnetically-driven winds with the scope of determining
the role and weight played by the various contributions in generating the
outflow. To this end, we consider a ``magnetized extension'' of the
classic Bernoulli criterion~\citep{Rezzolla_book:2013} that is
customarily used to determine whether an unmagnetized fluid element is
gravitationally bound to a gravitational field central object. More
specifically, assuming a stationary and axisymmetric spacetime, we
consider a magnetized fluid element unbound if~\citep{Bekenstein1978,
  Gourgoulhon2011}.
\begin{equation}
  \label{eq:Bern+}
  \left( h + \frac{b^2}{\rho} \right) u_t -\frac{b_t}{\kappa} < -1\,,
\end{equation}
where, again, $h$ is specific enthalpy of the fluid, $b$ and $b_t$ are
respectively the strength of the magnetic field in the comoving frame and
its covariant time component, $-u_t$ is the fluid energy at infinity (\ie
the covariant time component of the fluid four-velocity), and $\kappa :=
\rho u^{\rm pol}/b^{\rm pol}$ is proportional to the ratio of the
kinetic-to-magnetic energies in the comoving frame (in a spherical
coordinate system, $u^{\rm pol} := \sqrt{[(u^{\theta})^2 +
    (u^{r})^2]/2}$).

\begin{figure*}[h!]
  \center
  \includegraphics[width=0.8\textwidth]{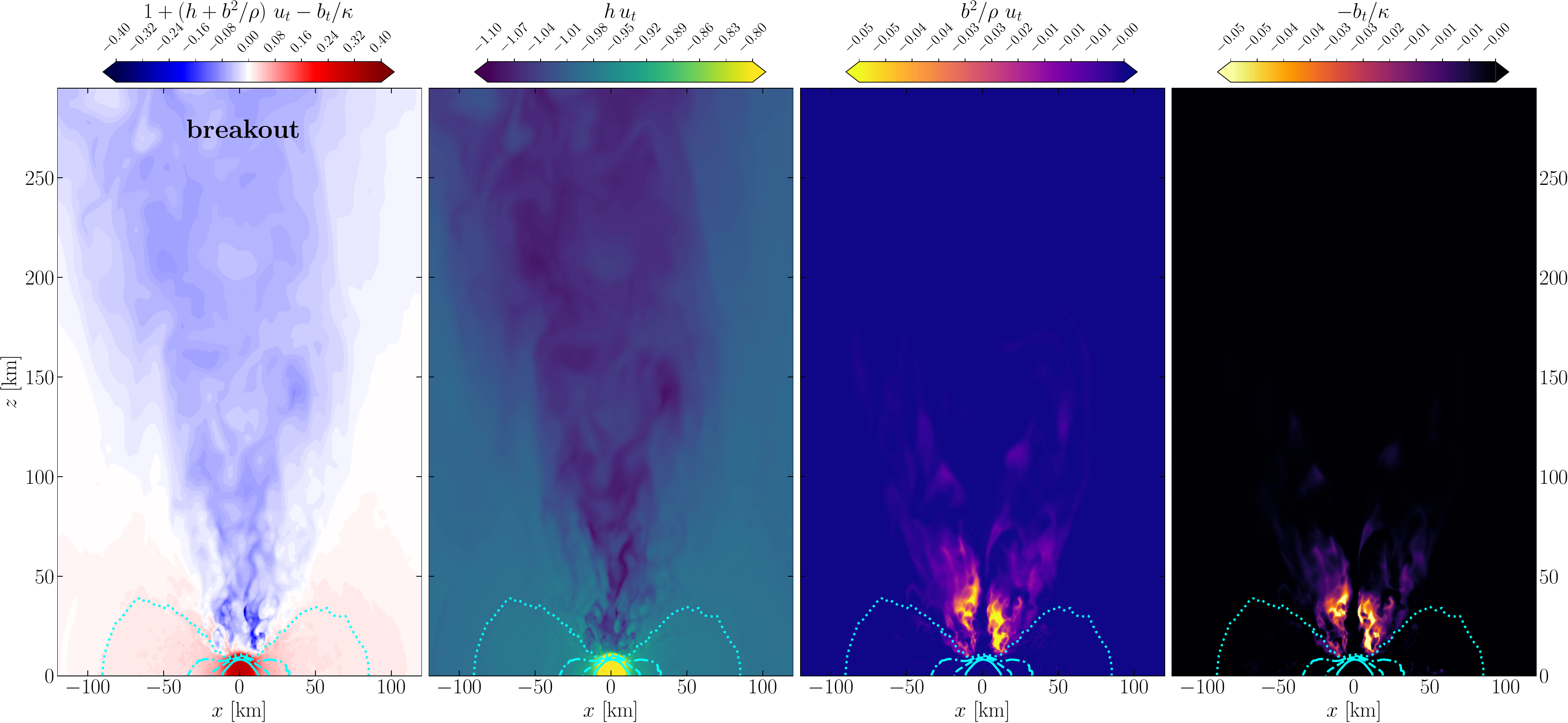}
  \vskip 0.25cm
  \includegraphics[width=0.8\textwidth]{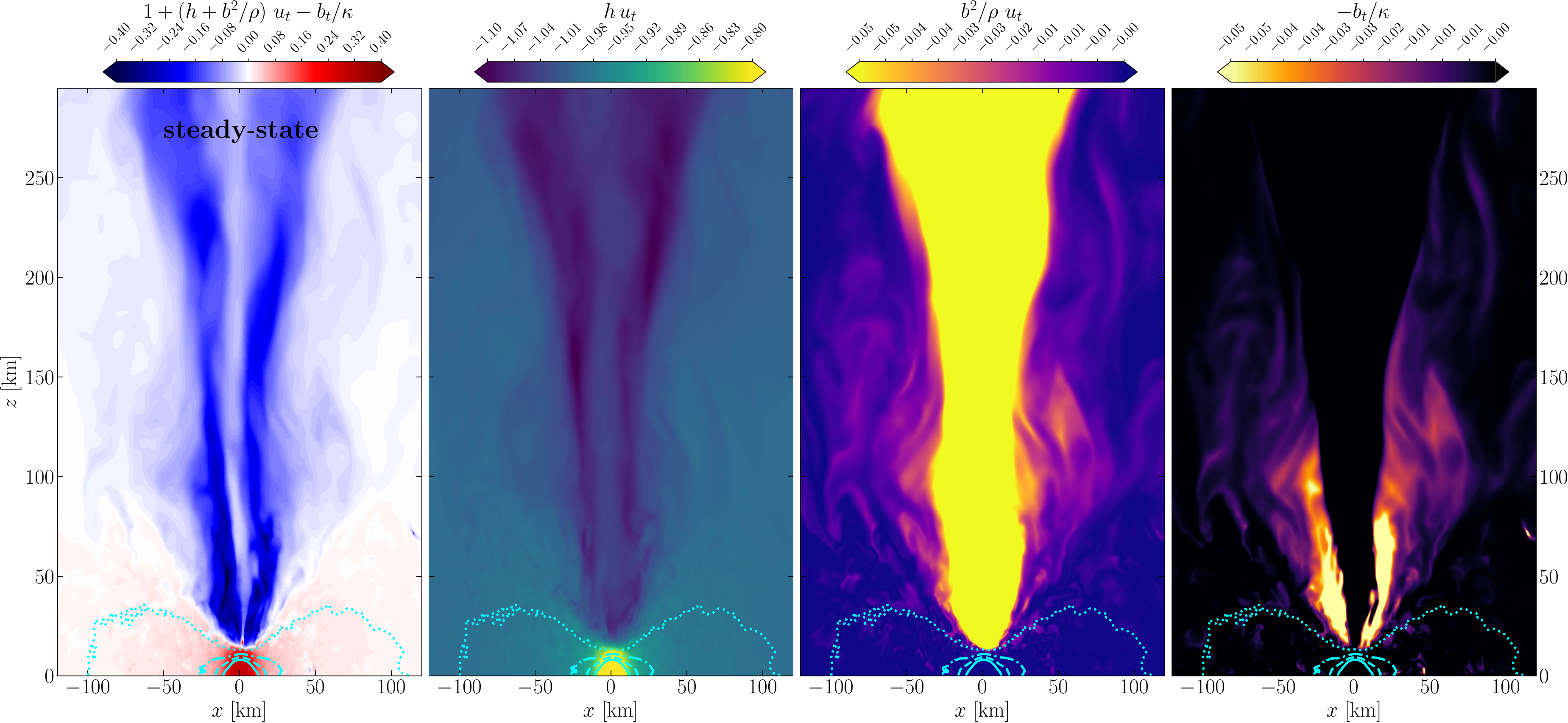}
  \caption{Two-dimensional distributions in the $(x,z)$ plane of various
    terms contributing to the extended Bernoulli criterion in
    Eq.~\eqref{eq:Bern+}. Shown from left to right are: the extended
    Bernoulli criterion $1+ ( h + b^2/\rho ) ~u_t - b_t/\kappa$, the
    geodesic criterion $h\,u_t$, the magnetization $b^2/\rho$, and the
    magneto-centrifugal contribution $b_t/\kappa$. The top row refers to
    the breakout time, while the bottom row to the steady-state stage;
    neutrinos are properly accounted in these simulations; see also
    Fig.~\ref{fig:pans_break}.}
  \label{fig:bernoulli_nu}
\end{figure*}

We can now analyse separately the various terms in
Eq.~\eqref{eq:Bern+}. Starting from the left, the first term $h\, u_t$ is
the classic and the purely kinetic ``geodesic'' term that is
traditionally considered when judging whether or not a fluid element is
bound or not~\citep[see, \eg][for a detailed discussion]{Bovard2016,
  Musolino2024}. The second term, on the other hand, is magnetic in
origin and is often referred to as the ``magnetisation'' of the plasma,
$\sigma:=b^2/\rho$, that is, the ratio of magnetic-to-rest-mass energy of
the fluid. This term accounts therefore for the purely magnetic energy
reservoir that can accelerate the plasma generating an outflow; the
larger $\sigma$, the larger the weight of the magnetic field and, of
course, $\sigma > 1$ signals a magnetically-dominated plasma. Finally,
the third term in~\eqref{eq:Bern+}, expresses the ratio between the
toroidal magnetic field and th so-called ``mass-loading'' parameter
$\kappa$ that is a combination of the poloidal components of the fluid
velocity and of the poloidal comoving magnetic field. As a result, in the
Newtonian limit the whole term $b_t/\kappa$ reduces to a magnetic
lever-arm contribution to the total fluid
energy~\citep{ZhuStone2018}\footnote{The Newtonian limit of
expression~\eqref{eq:Bern+} is given by $\Phi + \frac{1}{2} v^2 + h +
\frac{B^2}{2} + {\boldsymbol{B}\cdot
  \boldsymbol{v}}/{\kappa}$~\citep{ZhuStone2018}, where $\Phi$ is the
gravitational potential, while $\boldsymbol{B}$ and $\boldsymbol{v}$ are
the magnetic field and fluid-velocity vectors, respectively.}. This last
term, while only active in the outer layers of the wind, provides an
important contribution to the conserved energy of fluid elements in that
region, and is responsible for a Blandford-Payne type magneto-centrifugal
acceleration and collimation of the wind. Indeed, Blandford-Payne winds
in the context of Newtonian astrophysical simulations are identified
precisely through the evaluation of this energy component, which is
observed here for the first time in a general-relativistic
context. Clearly, the larger the third term in Eq.~\eqref{eq:Bern+}, the
larger the Blandford-Payne magneto-centrifugal acceleration imparted on
the fluid.

\begin{figure*}[h!]
  \center
  \includegraphics[width=0.8\textwidth]{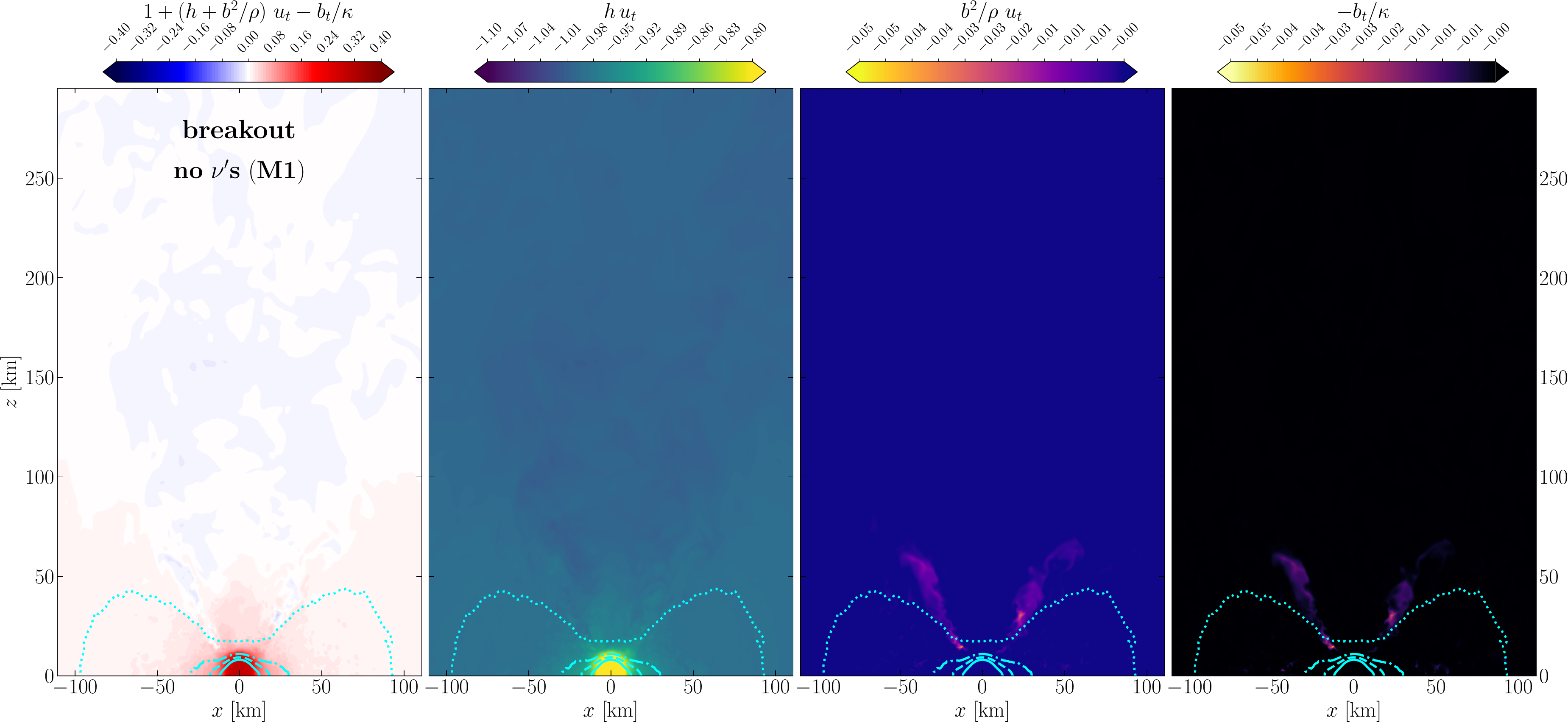}
  \vskip 0.25cm
  \includegraphics[width=0.8\textwidth]{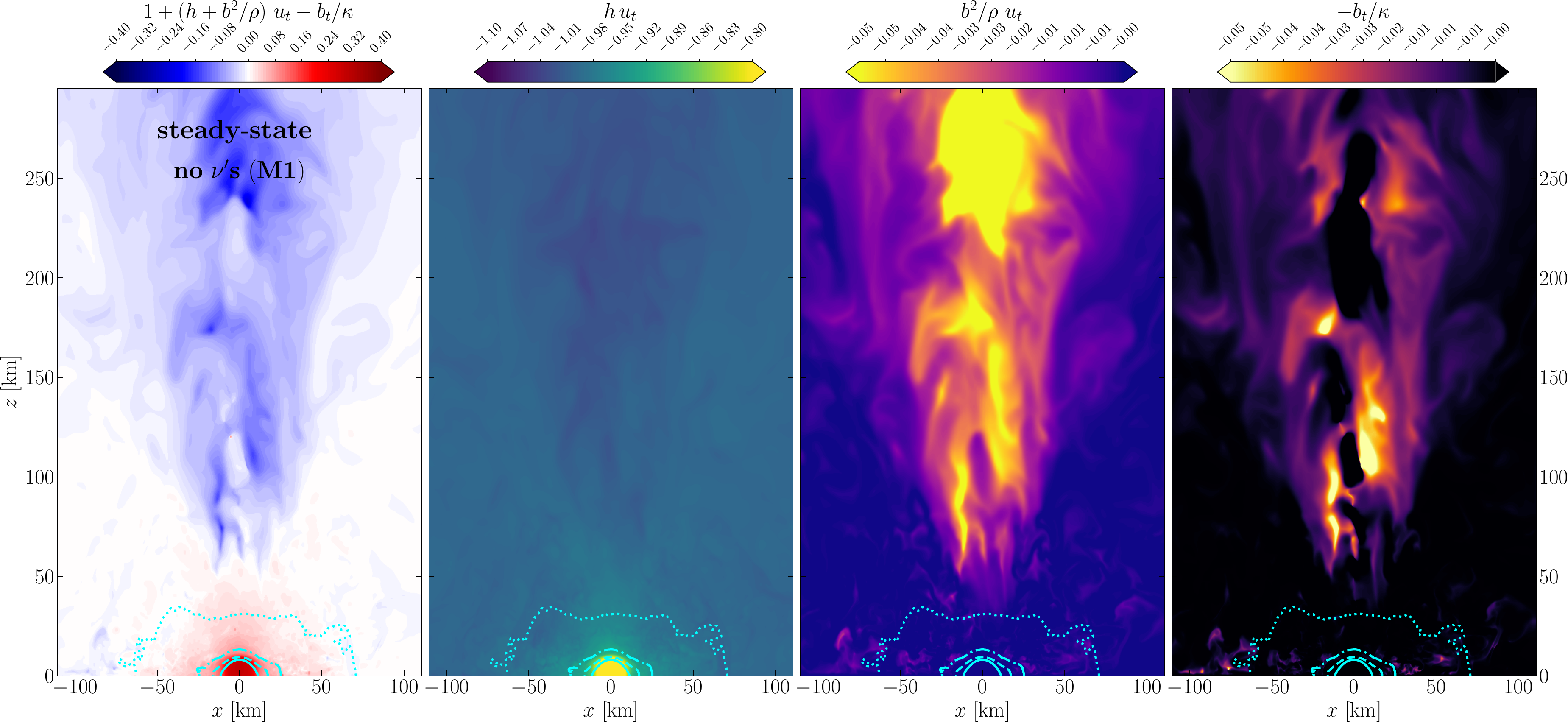}
  \caption{The same as in Fig.~\ref{fig:bernoulli_nu} but when neutrinos
    are not taken into account; see also Fig.~\ref{fig:pans_break}.}
  \label{fig:bernoulli_nonu}
\end{figure*}

Figure~\ref{fig:bernoulli_nu} provides two snapshots of all of these
terms at either the breakout time (top row) or in the steady-state phase
(bottom row). More specifically, from left to right, we present the
extended Bernoulli criterion $1+ ( h + b^2/\rho ) ~u_t - b_t/\kappa$, the
geodesic criterion $h\,u_t$, the magnetization $b^2/\rho$, and the
magneto-centrifugal contribution $b_t/\kappa$. When looking at the top
row, it is then apparent that at breakout only a mild wind is present
(first two panels from left) and that this wind is neutrino-driven, given
that the magnetization in the funnel is very small (third panel from
left; compare also with the first two panels in
Fig.~\ref{fig:bernoulli_nonu}) and is larger at the edges of the disk,
exactly where the $\alpha$-$\Omega$~dynamo will lead to the breakout of
the magnetic field. At this stage, the magneto-centrifugal acceleration
(rightmost panel) is clearly very small. On the other hand, exploring the
bottom row in the steady-state stage, it is clear that a strong unbound
wind is launched (first two panels from left) and that this is
magnetically-driven, given the very large magnetization in the funnel
(third panel from left). Importantly, note how at this point the
magneto-centrifugal acceleration provides a significant contribution and
that this is essentially absent in the regions very close to the
pole. This is simply because there is no lever-arm to produce a
magneto-centrifugal acceleration there, and this can be active only in
the upper edges of the disk. To the best of our knowledge, this is the
first time that a Blandford-Payne type of acceleration is shown to be
active in the metastable remnant from a BNS merger.

Figure~\ref{fig:bernoulli_nonu} provides the same type of diagnostic
shown in Fig.~\ref{fig:bernoulli_nu} but for the GRMHD simulations that
have neglected neutrino transport (no ${\rm M1}$). Contrasting the top
rows of the two figures, it is very simple (and obvious) to recognise
that no neutrino-driven wind is present in this latter case, but also
that no magneto-centrifugal acceleration can be observed. This provides a
very robust evidence that the outflowing wind measured before breakout is
generated by neutrinos. Similarly, when contrasting the bottom rows of
the two figures it is east to capture the very similar qualitative
behaviour of the outflows. However, as already commented for other fluid
quantities, the lack of neutrinos in Fig.~\ref{fig:bernoulli_nonu}
implies that the baryon pollution affects the outflow, which is therefore
less collimated, less magnetized and with a smaller Blandford-Payne
contribution. The disk too is impacted by the absence of neutrinos and of
the associated cooling. As a result, the disk will be more extended in on
the equatorial plane and also with a larger thickness (compare the top
rows of Figs.\ref{fig:bernoulli_nu} and \ref{fig:bernoulli_nonu}), making
the breakout harder to achieve. While our simulations end at about
$t-t_{\rm mer}\sim 110\,{\rm ms}$, it is reasonable to assume that that
the bottom panels of Figure~\ref{fig:bernoulli_nonu} will resemble
closely those of Fig.~\ref{fig:bernoulli_nu} on sufficiently large
timescales.

\begin{figure*}[h!]
  \center
  \includegraphics[width=0.8\textwidth]{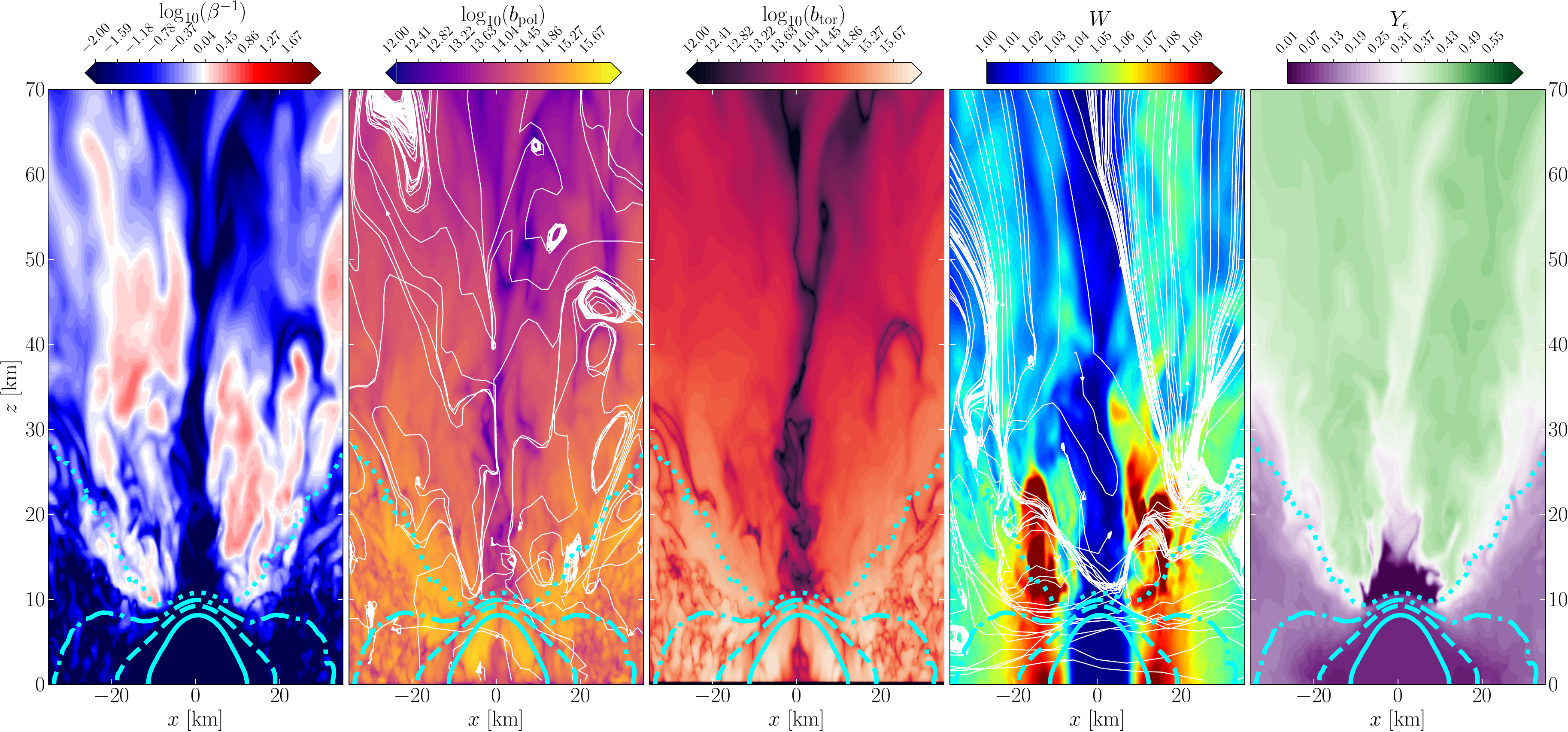}
  \vskip 0.25cm
  \includegraphics[width=0.8\textwidth]{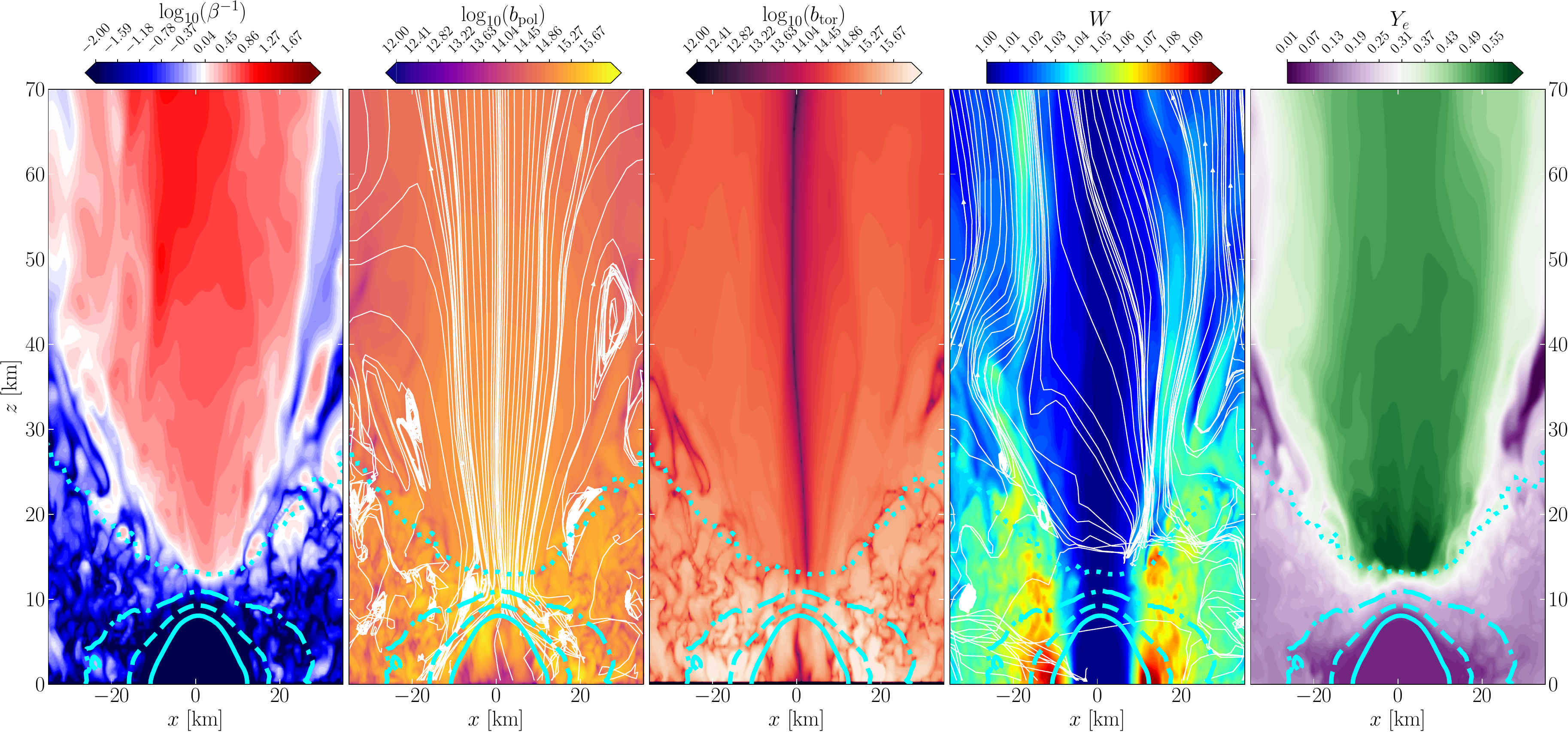}
  \caption{The same as Fig.~\ref{fig:pans_break} but on a smaller scale to highlight
    features near the HMNS.}
  \label{fig:pans_break_m1_zoom}
\end{figure*}

\section{Close-up view of the plasma properties}
\label{sec:app_C}

We have already commented on the features of Fig.~\ref{fig:pans_break}
and how the two rows illustrate the status of various properties of the
plasma at breakout (top bottom rows and in the steady-state stage (bottom
row). Figure~\ref{fig:pans_break_m1_zoom} provides the same information
but when zooming-in on lengthscales that are representative of the
HMNS. In this case, a number of features can be appreciated at breakout:
(i) the breakout clearly takes place from the disk and not from the HMNS
and, in particular, from plasma at densities between $10^{11}$ and
$10^{12}\,{\rm G}$; (ii) the toroidal magnetic field is weak along the
rotation axis and essentially zero in the core of the HMNS; (iii) because
of the high differential rotation of the HMNS, the maximum velocity and
hence Lorentz factors are achieved at $\simeq 10$-$15\,{\rm km}$ from the
rotation axis and are very small on the rotation axis because of the
axial symmetry; (iv) the regions where neutrinos start to modify the
electron fraction coincide with those where the breakout takes place,
leaving the regions very close to the pole of the HMNS with very small
values of $Y_e$. On the other hand, in the late steady-state stage we
note that: (i) the magnetically-driven wind now comes also from the polar
regions of the HMNS, filling the funnel completely; (ii) the regions
where the toroidal magnetic field is weak reduce considerably and are
only along the rotation axis; (iii) because of high-speed magnetically
driven wind, the regions where breakout took place, \ie on the disk
edges, have much lower velocities now in the azimuthal direction, while
differential rotation in the HMNS does not change considerably; (iv) the
regions very close to the pole of the HMNS have now larger values of
$Y_e$ and although this is true for the whole funnel, rapid changes still
take place when moving from the center to the pole of the HMNS.

\section*{Acknowledgements}

The authors acknowledges insightful discussions with L. Combi,
O. Gottlieb, K. Kiuchi, B. Metzger, P. M\"osta, D. Siegel and
A. Tchekhovskoy. Partial funding comes from the State of Hesse within the
Research Cluster ELEMENTS (Project ID 500/10.006), by the ERC Advanced
Grant ``JETSET: Launching, propagation and emission of relativistic jets
from binary mergers and across mass scales'' (Grant No. 884631). CE
acknowledges support by the Deutsche Forschungsgemeinschaft (DFG, German
Research Foundation) through the CRC-TR 211 ``Strong-interaction matter
under extreme conditions''-- project number 315477589 -- TRR 211. LR
acknowledges the Walter Greiner Gesellschaft zur F\"orderung der
physikalischen Grundlagenforschung e.V. through the Carl W. Fueck
Laureatus Chair. ERM is supported by the National Science Foundation
under grant No. PHY-2309210. The calculations were performed in part on
the local ITP Supercomputing Clusters Iboga and Calea and in part on HPE
Apollo HAWK at the High Performance Computing Center Stuttgart (HLRS)
under the grants BNSMIC and BBHDISKS.

\section*{Data Availability}
Data is available upon reasonable request from the corresponding
author.

\end{document}